\documentclass[fleqn,usenatbib, useAMS]{mnras}

\usepackage{newtxtext,newtxmath}

\usepackage[T1]{fontenc}
\usepackage{ae,aecompl}
\usepackage{graphicx}	
\usepackage{times}

\title[Stellar splashback]{Stellar splashback: the edge of the intracluster light}

\author[A. J. Deason et al.]{
Alis J. Deason$^{1,2}$\thanks{E-mail: alis.j.deason@durham.ac.uk},
Kyle A. Oman$^{1}$, Azadeh Fattahi$^{1}$, Matthieu Schaller$^{3}$, \newauthor Mathilde Jauzac$^{2,1,4,5}$, Yuanyuan Zhang$^{6}$, Mireia Montes$^{7}$, Yannick M. Bah\'e$^{3}$, \newauthor Claudio Dalla Vecchia$^{8,9}$, Scott T. Kay$^{10}$, Tilly A. Evans$^{1,2}$
\\
$^{1}$Institute for Computational Cosmology, Department of Physics, University of Durham, South Road, Durham DH1 3LE, UK\\
$^{2}$Centre for Extragalactic Astronomy, Department of Physics, University of Durham, South Road, Durham DH1 3LE, UK\\
$^{3}$ Leiden Observatory, Leiden University, PO Box 9513, NL-2300 RA Leiden, The Netherlands\\
$^{4}$Astrophysics Research Centre, University of KwaZulu-Natal, Westville Campus, Durban 4041, South Africa\\
$^{5}$School of Mathematics, Statistics \& Computer Science, University of KwaZulu-Natal, Westville Campus, Durban 4041, South Africa\\
$^{6}$Fermi National Accelerator Laboratory, P.O. Box 500, Batavia, IL 60510, USA\\
$^{7}$School of Physics, University of New South Wales, Sydney, NSW 2052, Australia\\
$^{8}$Instituto de Astrof\'isica de Canarias, C/V\'ia L\'actea s/n, E-38205 La Laguna, Tenerife, Spain\\
$^{9}$Departamento de Astrof\'isica, Universidad de La Laguna, Av.~Astrof\'isico Francisco S\'anchez s/n, E-38206 La Laguna, Tenerife, Spain\\
$^{10}$ Jodrell Bank Centre for Astrophysics, Department of Physics and Astronomy, School of Natural Sciences, The University of Manchester, Manchester M13 9PL, UK
}
\date{Accepted XXX. Received YYY; in original form ZZZ}

\pubyear{2020}

\begin{document}
\label{firstpage}
\pagerange{\pageref{firstpage}--\pageref{lastpage}}
\maketitle

\begin{abstract}
  We examine the outskirts of galaxy clusters in the C-EAGLE simulations to quantify the `edges' of the stellar and dark matter distribution. The radius of the steepest slope in the dark matter, commonly used as a proxy for the splashback radius, is located at $\sim \!  \! r_{200 \rm m}$; the strength and location of this feature depends on the recent mass accretion rate, in good agreement with previous work. Interestingly, the stellar distribution (or intracluster light, ICL) also has a well-defined edge, which is directly related to the splashback radius of the halo. Thus, detecting the edge of the ICL can provide an independent measure of the physical boundary of the halo, and the recent mass accretion rate. We show that these caustics can also be seen in the projected density profiles, but care must be taken to account for the influence of substructures and other non-diffuse material, which can bias and/or weaken the signal of the steepest slope. This is particularly important for the stellar material, which has a higher fraction bound in subhaloes than the dark matter. Finally, we show that the `stellar splashback' feature is located beyond current observational constraints on the ICL, but these large projected distances ($\gg 1$ Mpc) and low surface brightnesses ($\mu \gg 32$ mag arcsec$^{-2}$) can be reached with upcoming observational facilities such as the Vera C. Rubin Observatory, the Nancy Grace Roman Space Telescope, and Euclid.
 \end{abstract}


\section{Introduction}
The dark matter haloes that underpin our hierarchical structure formation paradigm do not have uniquely defined boundaries. Several common definitions are used in the literature: the `friends-of-friends' distance \citep{davis85}, the virial radius \citep{bryan98}, and a radius within which the mean density equals a fixed value times the critical density or the cosmic mean value (spherical overdensity halo boundaries\footnote{We adopt the common notation for subscripts, $\Delta_{\rm c}$ or $\Delta_{\rm m}$, where $\Delta$ represents the overdensity with respect to the critical density (c) or the cosmic mean (m) matter density, e.g. $r_{200\rm c}$, $r_{200 \rm m}$.}).  Regardless of the exact definition, the use of a halo boundary is essential in order to define halo masses, distinguish between field and satellite galaxies, and, importantly, contrast the predictions of simulations with observations. Often the choice of halo boundary depends on the mass scale under consideration, and whether the study is observationally or theoretically motivated. A common definition across halo mass and redshift that is also observationally motivated (or even applicable) is crucial.

Recent work has argued that the most physical definition of the halo boundary is related to the transition between collapsed and infalling material, or the one- and two-halo regimes, and has been termed the `splashback' radius \citep[e.g][]{adhikari14, diemer14, more15, diemer20}. The splashback radius corresponds to the first apocentre of recently infalling dark matter; \cite{diemer14} showed that this radius can often be identified in cosmological simulations from the radius of steepest slope in the density profiles of the dark matter,  where a sharp drop is caused by particles piling up at their apocentre. This splashback radius does not only define a physical halo boundary, it also crucially depends on the mass accretion rate of the collapsing halo \citep[e.g][]{adhikari14, diemer14, diemer17}. Importantly, there is now considerable evidence that the splashback radius has been identified in galaxy clusters, either through stacked satellite galaxy surface density profiles \citep{more16, baxter17, shin19, zurcher19, murata20}, or weak lensing \citep{chang18, contigiani19, tam20}.
Initially, the location of the observed splashback radius appeared to be lower (by $\sim  \! \! 20\%$) than the predictions of $\Lambda$CDM simulations \citep[e.g][]{more16,baxter17, chang18}. However, it has since be convincingly shown that this is a result of selection effects from the optical cluster finding algorithms \citep[e.g.][]{busch17, murata20, xhakaj20}. Accurate observational measurements are vital as any discrepancies with the $\Lambda$ cold dark matter ($\Lambda$CDM) predictions could signal more exotic solutions, such as self-interacting dark matter \citep{more16, banerjee20}.

The theoretical background to spherical overdensity halo boundaries, and the more recently promoted splashback radius, is based almost entirely on the dark matter distribution. This is perhaps unsurprising, as the outer reaches of galaxy and cluster haloes are dominated by dark matter, and the majority of the visible material is concentrated at the very centre. However, the use of an observationally motivated halo boundary, defined using the baryonic material, is, in some cases, more attractive. Satellite galaxies are an obvious way forward: they can be identified with photometric and spectroscopic surveys, and can reach out to large distances in galaxy haloes \citep[e.g][]{vandenbosch05, yang05, robotham11, budzynski12, mcconnachie12}. The main drawback is that the number of visible satellite galaxies can be low for individual systems, and care must be taken to understand the selection effect of a stellar mass limited sample \citep[see e.g.][]{adhikari16} and the impact of satellite galaxy colour \citep[e.g.][]{baxter17, shin19, adhikari20}. An obvious, complementary, probe of the halo is the remains of \textit{destroyed} satellite galaxies (i.e. the stellar halo), which are commonly referred to as the intracluster light (ICL) at cluster scales \citep[e.g.][]{mihos16, montes19_rev}. 

Recently, \cite{deason20} provided the first foray into defining the \textit{stellar} edges of galactic haloes (with halo mass $\sim 10^{12}\mathrm{M}_\odot$). These authors used  high-resolution cosmological hydrodynamic simulations of Milky Way-mass haloes to explore the boundary of the halo stars. Curiously, they found that the stars have a well-defined edge, but this is not coincident with the splashback radius of the dark matter. Rather, the edge of the halo stars appears to be related to a secondary dark matter caustic (termed the `second caustic' in this work), which likely corresponds to the edge of the virialized material that has completed at least two pericentric passages. However, extrapolating these findings to other mass scales is non-trivial owing to the non-linear stellar mass--halo mass relation \citep{moster10, behroozi13} and the varying importance of `smooth' accretion onto galaxy haloes with halo mass \citep[e.g][]{genel10, fakhouri10}. 

The stellar haloes of Milky Way-mass galaxies are primarily built from the leftover debris from destroyed dwarf galaxies \citep[e.g][]{bullock05, cooper10, deason15,deason16}. On cluster-mass scales, the ICL is built predominantly from the destroyed remnants of Milky Way-mass galaxies \citep[e.g.][]{murante04, conroy07, purcell07, puchwein10, contini14, montes14, montes18, demaio18}. This self-similarity from dwarf galaxies to Milky Way-mass galaxies to clusters is a beautiful example of hierarchical structure formation in action. Nonetheless, the detailed properties of galactic stellar haloes and the ICL have important differences, most notably the significance of this component to the total stellar mass, and their radial distributions \citep[e.g.][]{pillepich14, pillepich18}. Indeed, while both galactic stellar haloes and the ICL form via mergers, important galaxy formation physics underpins their differences and similarities. Thus, studying these diffuse halo components over a range of mass scales allows for a critical view on both structure formation and models of galaxy formation.

In this work, as a complement to the \cite{deason20} study, we focus on the stellar haloes of clusters, i.e., the ICL. Recent work has shown an intriguing similarity between the dark matter density profiles of clusters and their ICL \citep[e.g.][]{pillepich18, montes19, asensio20}. This finding particularly motivates an investigation into the stellar edges of cluster-mass haloes. Here, we use the Cluster-EAGLE (C-EAGLE) suite of simulations to study the outer density profiles of both stars and dark matter, and their relation to each other. The paper is arranged as follows. In Section \ref{sec:ceagle}, we describe the C-EAGLE simulations, and in Section \ref{sec:edges}, we probe the edges of these galaxy clusters using both stars and dark matter. We explore the observationally motivated projected density profiles in Section \ref{sec:icl}, and discuss the implications for current and future observational probes of the ICL. Finally, we summarize our main conclusions in Section \ref{sec:conc}.

\section{C-EAGLE Simulations}
\label{sec:ceagle}
\begin{figure*}
\begin{minipage}{0.45\textwidth}
        \includegraphics[width=\textwidth,angle=0]{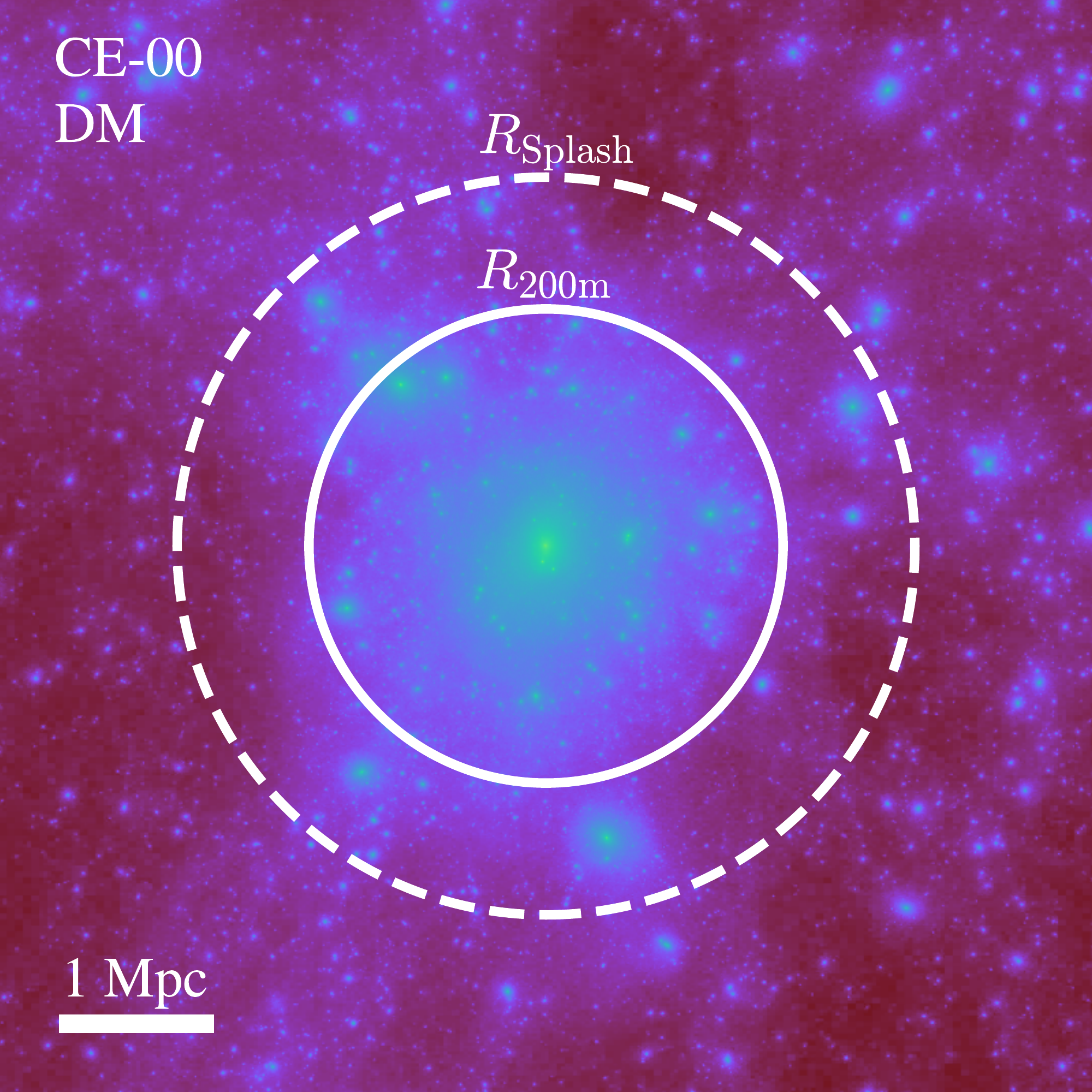}
 \end{minipage}       
 \begin{minipage}{0.45\textwidth}
        \includegraphics[width=\textwidth,angle=0]{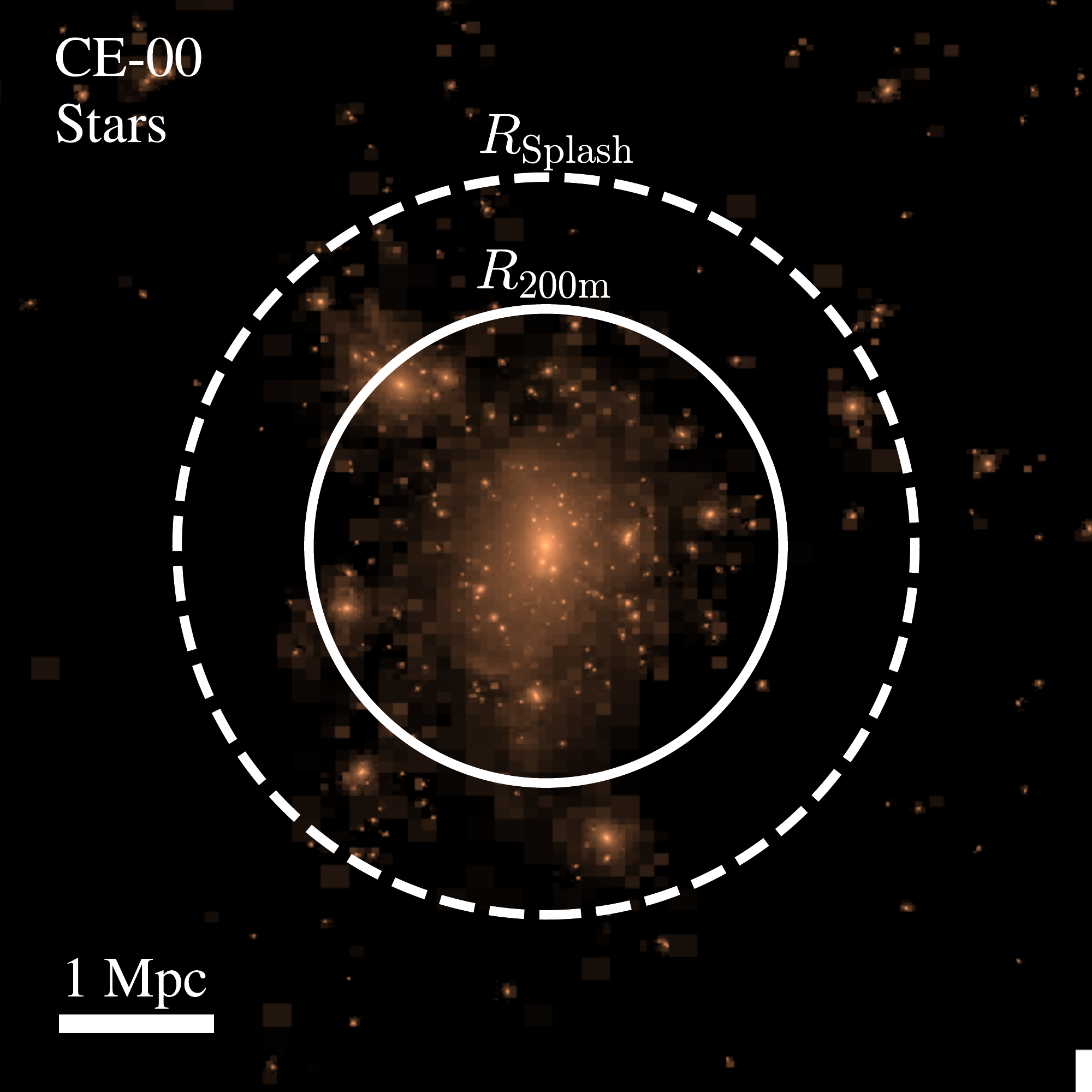}
 \end{minipage}     
 \begin{minipage}{0.45\textwidth}
        \includegraphics[width=\textwidth,angle=0]{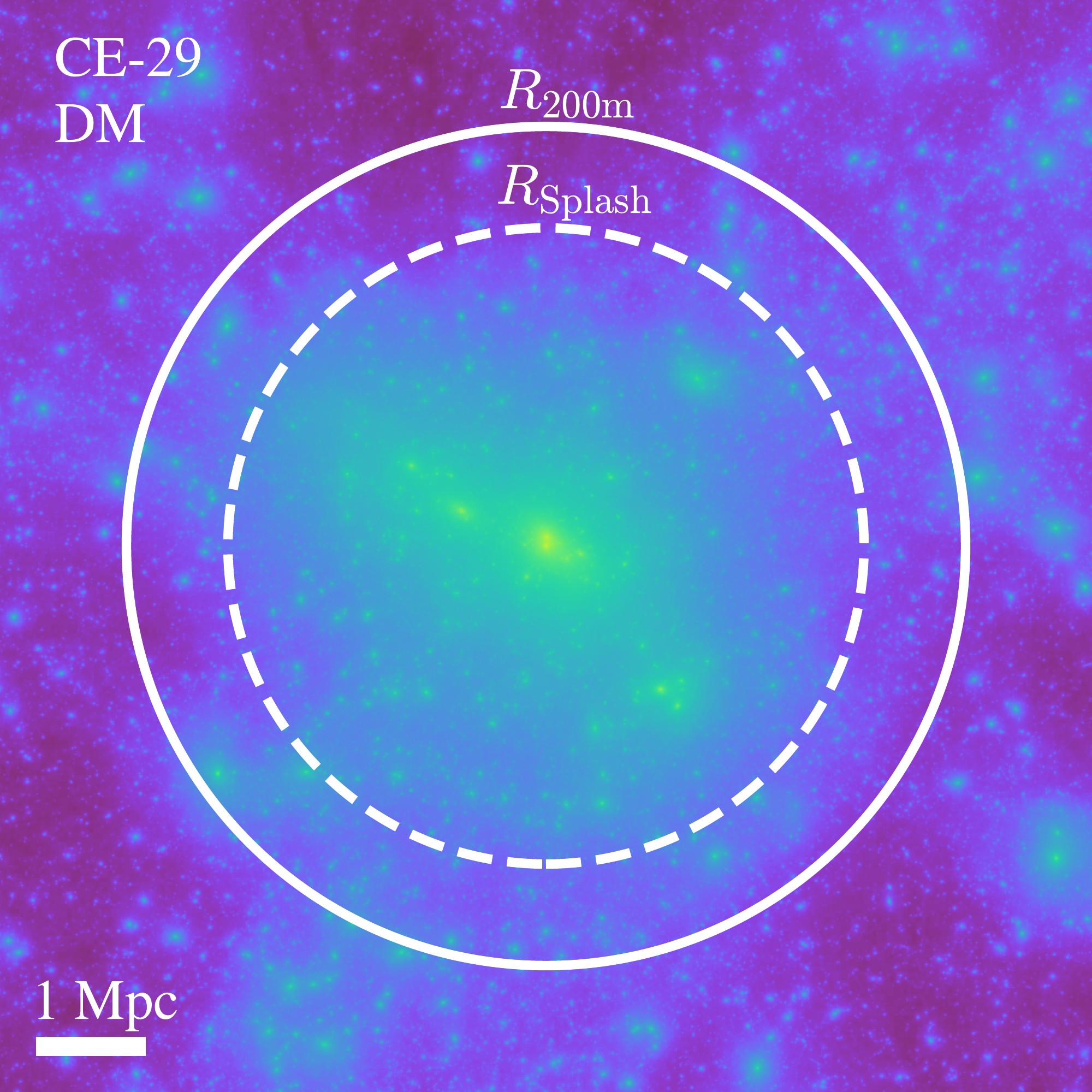}
 \end{minipage}       
 \begin{minipage}{0.45\textwidth}
        \includegraphics[width=\textwidth,angle=0]{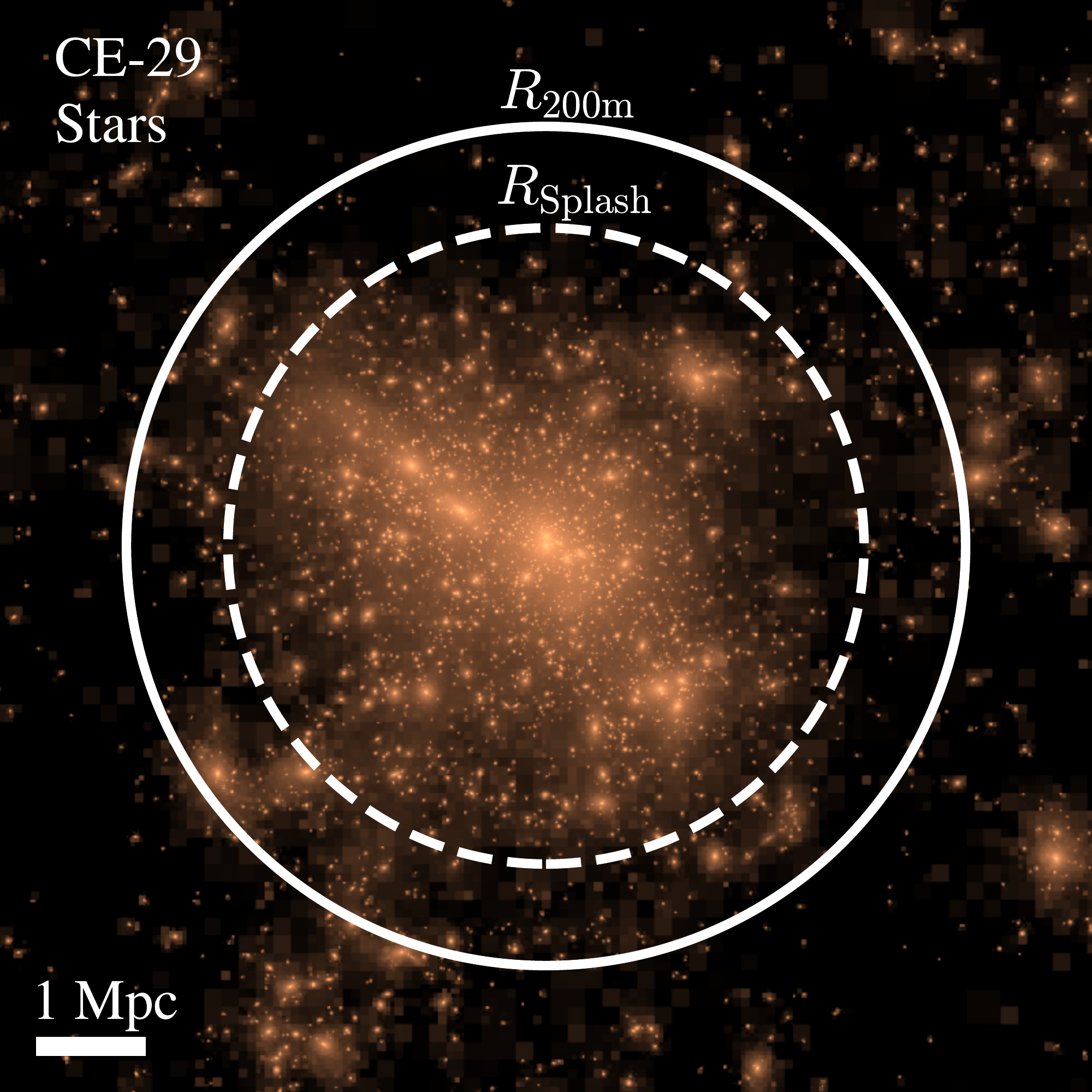}
 \end{minipage}  
         \caption[]{Projection of dark matter (left hand panels) and stars (right hand panels) for two example haloes at $z=0$ (CE-00: $M_{\rm 200 m} = 1.7 \times 10^{14}M_\odot$; and CE-29: $M_{\rm 200 m}= 3.2 \times 10^{15}M_\odot$). The solid line indicates the spherical overdensity boundary $r_{200\rm m}$, and the dashed line shows the splashback radius (see Section \ref{sec:edges}). The color scale is logarithmic, with projected density values ranging from $1\times 10^{-1}$ to $1\times 10^4$ and $1 \times 10^{-3}$ to $3\times 10^2$ M$_\odot$/pc$^2$ for the dark matter and stellar distributions, respectively. This image was produced using the open source project \texttt{yt} \citep{yt}.}
          \label{fig:images}
\end{figure*}

In this work, we use the C-EAGLE project \citep{bahe17,barnes17} to study the outer density profiles of galaxy clusters. This suite is a set of $N=30$ zoom-in cosmological hydrodynamical simulations of galaxy clusters in the mass range $10^{14.0} < M_{\rm 200 c}/\mathrm{M}_\odot < 10^{15.4}$. The simulations are run with the EAGLE galaxy formation model (AGNdT9 calibration, \citealt{schaye15}), with a gas particle mass of $1.8 \times 10^6\mathrm{M}_\odot$, a dark matter particle mass of $9.7 \times 10^6 \mathrm{M}_\odot$, and a physical softening at $z < 2.8$ of $0.7$ kpc. The clusters are selected from the parent low-resolution volume described in \cite{barnes17b}. The zoom-in technique isolates the selected clusters, and re-simulates the cluster region and its immediate environment at higher resolution. This ensures the area of interest is computed with high resolution, while the long range forces of gravity are still captured in their appropriate cosmological context.
 The high-resolution volumes are set up such that they are devoid of any low-resolution particles within at least $5r_{200 \rm c}$, and the clusters were selected to have no massive neighbours within $10r_{200 \rm c}$. Here, $r_{200 \rm c}$ is the radius at which the average density drops to 200 times the critical density at $z=0$.  A subset (24) of the C-EAGLE sample has been simulated at high resolution out to at least $10r_{200 \rm c}$; these are called the \textit{Hydrangea} simulations \citep{bahe17}. The simulations assume a flat $\Lambda$CDM cosmology with parameters \citep{planck14}:  $\Omega_{\rm m} = 0.307$, $\Omega_{\rm b} = 0.04825$, $\Omega_\Lambda = 0.693$, $h = 0.6777$, $\sigma_8 = 0.8288$ and $n_s = 0.9611$.

The EAGLE model is described in detail in \cite{schaye15} and \cite{crain15}, and includes subgrid models for baryonic processes such as star formation, stellar winds, gas cooling, metal production and stellar and black hole feedback.
These subgrid recipes were calibrated to reproduce the present-day stellar mass function, the galaxy size--stellar mass relation and the black hole mass--host galaxy mass relation. Note that since the EAGLE model was calibrated on galaxy properties, and not specifically on clusters, the properties of the C-EAGLE cluster sample are predictions of a model that produces realistic galaxies in the field. Projected images for two example clusters are shown in Fig. \ref{fig:images}. Here, we show the dark matter (left hand panels) and stellar mass (right hand panels) distributions.

The low-redshift global properties of the C-EAGLE sample are described in \citet[][see also \citealt{bahe17}]{barnes17}. These works showed that the total stellar content, metal content (see also \citealt{pearce20}) and black hole masses are in good agreement with the observations. However, the clusters are too gas rich, their central temperatures are too high, and they have larger entropy cores than observed. These mismatches with observations are likely driven by shortcomings in the AGN feedback model. Of relevance to this work, \cite{asensio20} recently studied the ICL of the C-EAGLE clusters and found that the shape of the stellar mass distribution closely follows that of the total matter, in good agreement with observations \citep{montes19}. Moreover, Bah\'e et al. (in preparation) also find that the ICL surface density profiles agree with observations. In this work, we focus on the `edges' of these clusters and compare the stellar and dark matter halo boundaries.

\section{The Edge of Galaxy Clusters}
\label{sec:edges}
\begin{figure*}
 \centering
        \includegraphics[width=\textwidth,angle=0]{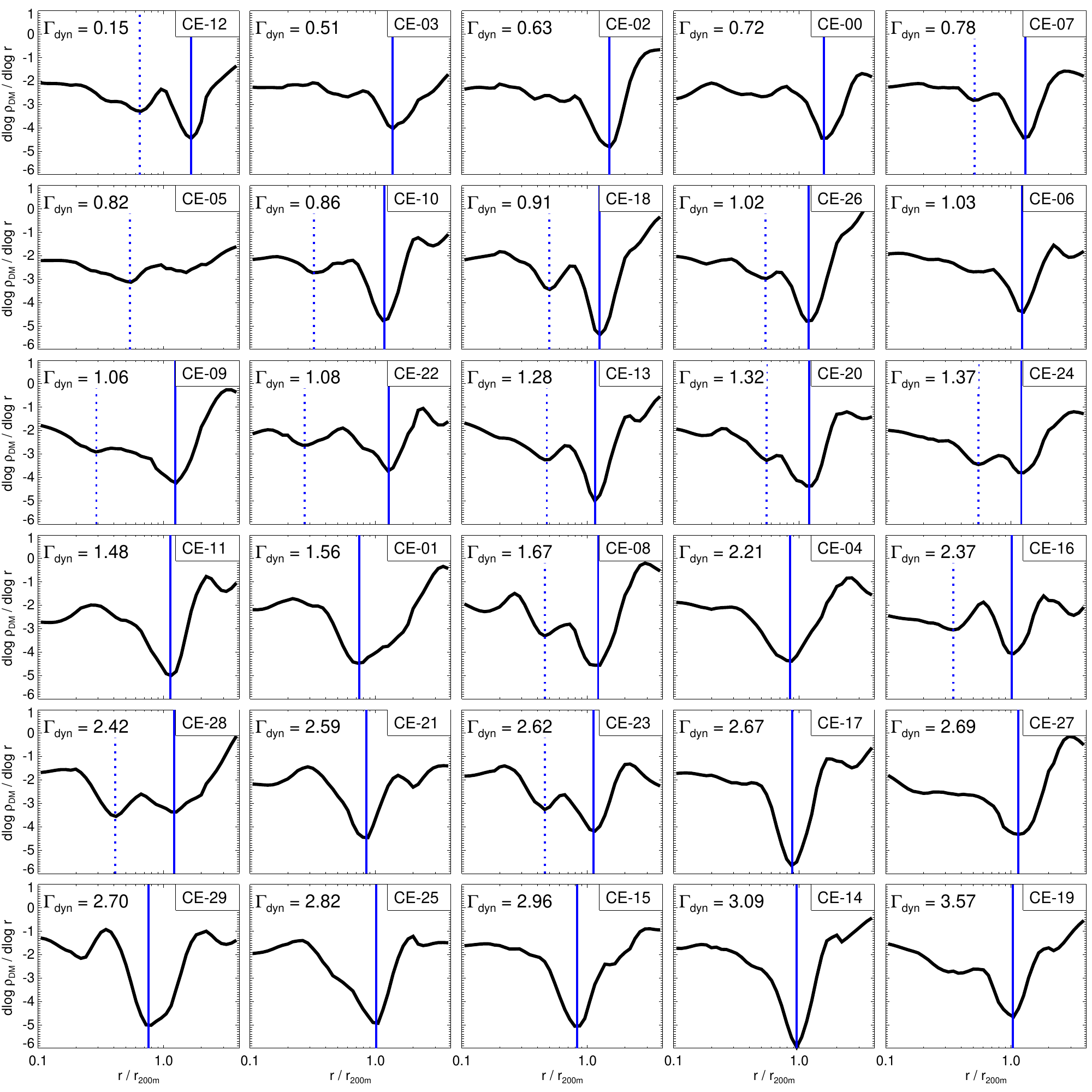}
         \caption[]{The logarithmic slope profiles, $\mathrm{d\,log}(\rho)/\mathrm{d\,log}(r)$, of the dark matter density profiles for the $N=30$ C-EAGLE clusters.  The profiles for individual clusters are computed using the angular median method (see the main text), and 40 evenly spaced radial bins have been used in the range $\mathrm{log}(r/r_{\rm 200m}) \in [-1.0, 0.6]$. The logarithmic profile is computed using the fourth-order Savitzky–-Golay smoothing algorithm over the 15 nearest bins  \citep{savitzky64}. The clusters are ordered according to the recent mass accretion rate, $\Gamma_{\rm dyn}$, increasing from the top left panel to the bottom right panel. The cluster ID is also indicated (see \citealt{barnes17}, table A1). The solid vertical lines show the most prominent minimum, defined as $r_{\rm Caustic}$, or the splashback radius. We also show with the dotted lines cases with clear second caustics. These are much weaker than the splashback radii, and tend to be more common amongst haloes with low recent mass accretion rates.}
          \label{fig:dm_slopes}
\end{figure*}

\begin{figure*}
 \centering
        \includegraphics[width=\textwidth,angle=0]{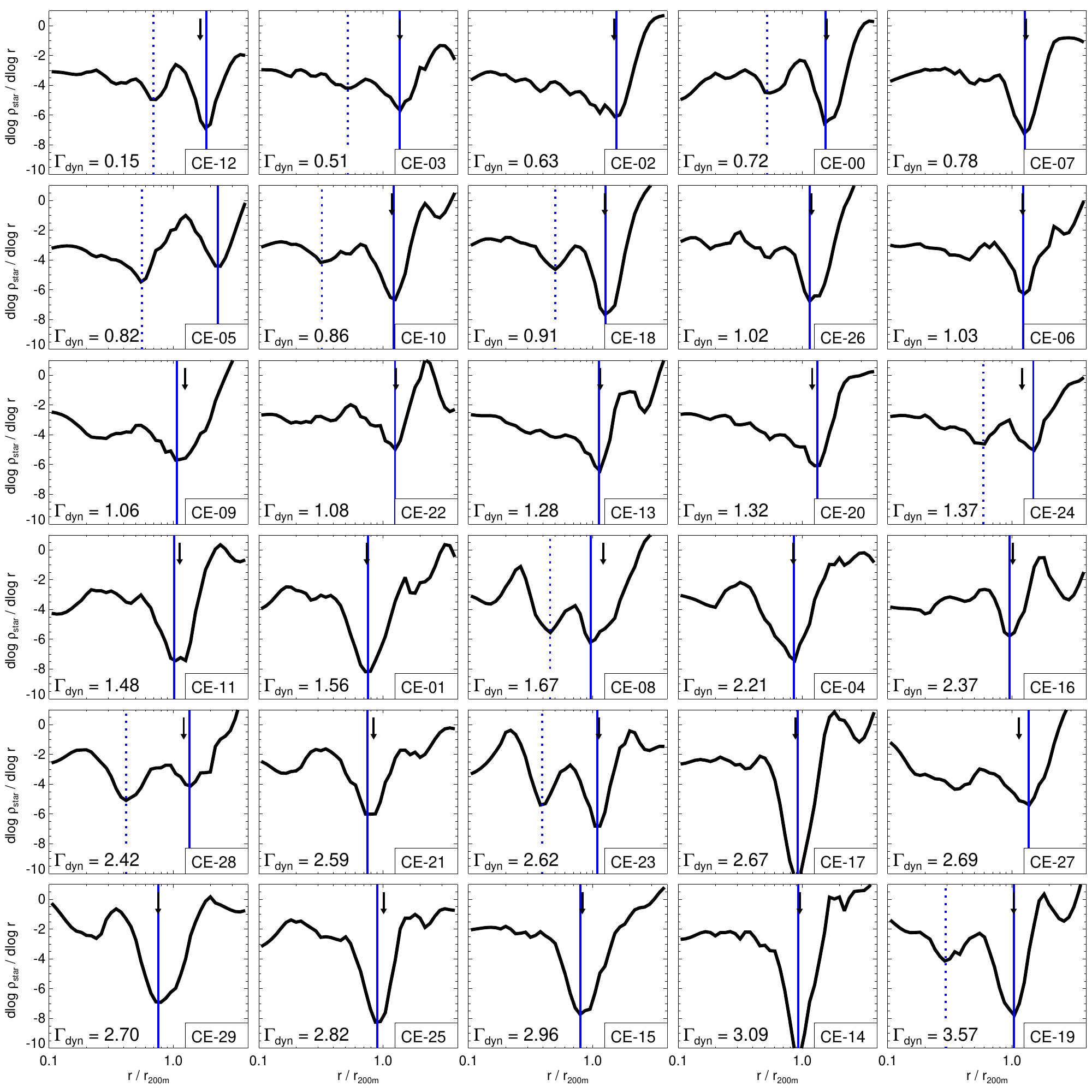}
         \caption[]{Same as Figure \ref{fig:dm_slopes}, but for the stellar density profiles of the C-EAGLE clusters (note the different $y$-axis scale). The black arrows indicate the dark matter splashback radius.}
          \label{fig:star_slopes}
\end{figure*}

In this section, we probe the dark matter and stellar density profiles of the C-EAGLE clusters. Following the work by \cite{diemer14}, we use the differential logarithmic density profiles to identify the steepest slope, which signifies a transition between the collapsed (one-halo) and infalling (two-halo) material. Throughout this work, we consider the radius of steepest slope as a proxy for the splashback radius. We consider all\footnote{This includes \textit{all} particles in the halo, not just the particles identified by \textsc{subfind} to be in the main subhalo. So, particles in subhaloes and unbound particles are also included.} dark matter and stellar particles in the simulations out to a 3D radius of $4r_{200\rm m}$ from the halo centre. Here, $r_{200 \rm m}$ is the radius at which the average density drops to 200 times the universal matter density at $z=0$. Throughout this work, we scale physical radii with this radius. Our outer boundary of $4r_{200\rm m}$ sometimes contains a small fraction of low-resolution dark matter particles. However, this makes little difference to our results as we are mainly interested in the region within $\sim \! \! 2r_{200 \rm m}$, which is completely devoid of any low-resolution particles. Note, for ease of comparison, $r_{200 \rm m} \sim \! 1.7 r_{200 \rm c}$ at the cluster mass scale.

For both dark matter and stars we construct density profiles in 40 evenly spaced logarithmic bins between $0.1$ and $4$ $r/r_{200\rm m}$. We follow a similar approach to \cite[][see their section 4.3]{mansfield17} in order to construct the angular median density profile. Namely, for each logarithmic radial shell, the density profile is computed in $N=50$ (equally spaced) solid angle segments. We construct
the density profile by taking the median of these profiles in each radial shell. This procedure minimizes the influence of massive substructures and other non-diffuse structures on the density profile. As we will show in Section \ref{sec:icl}, the median angular profile is far more effective at isolating the steepest halo slope than the more commonly used mean; this is particularly important for the stellar distribution. We note that the number of angular bins in this procedure is chosen as a balance between accounting for the effect of outliers, and ensuring that our results are not badly affected by noise. Our fiducial number is $N=50$ (the same as \citealt{mansfield17}), however, when this number is varied by a factor of 2 (i.e. $25$ or $100$ angular bins), our estimated splashback radii are changed by less than 10\%. Finally, we compute the logarithmic slope profiles using a fourth-order Savitzky–-Golay smoothing algorithm \citep{savitzky64} over the 15 nearest bins. This bin size and smoothing was chosen to minimize noise, while allowing us to identify the strongest features in the profile.

\begin{figure*}
 \centering
        \includegraphics[width=\textwidth,angle=0]{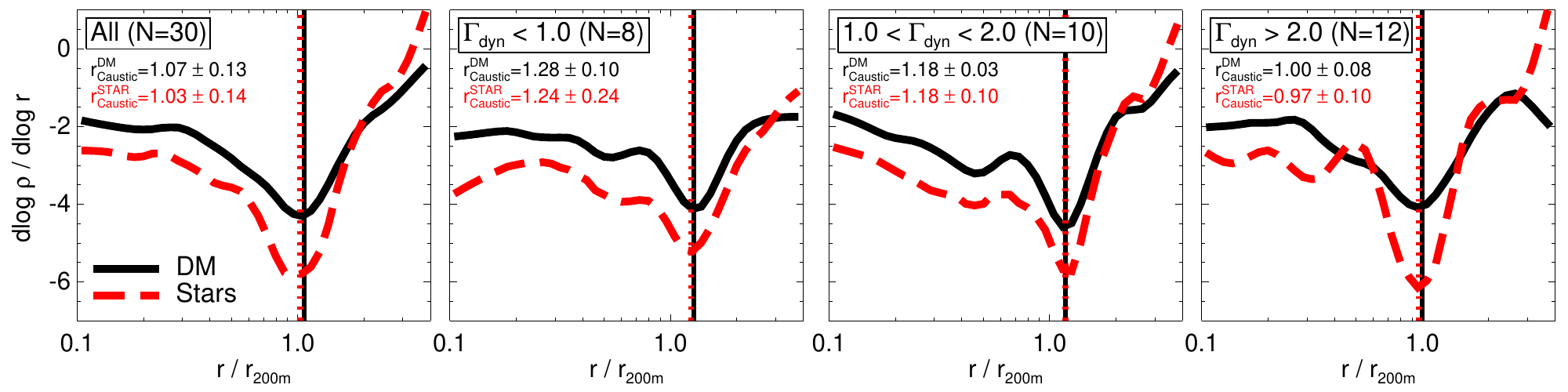}
         \caption[]{The stacked logarithmic slope profiles of the dark matter (solid black) and stellar (dashed red) density profiles of the C-EAGLE clusters. Each individual profile is determined using the angular median method and we show the median over all clusters. In the left-hand panel the median of all $N=30$ clusters is shown. In the remaining panels, we show the median profiles for relatively low ($\Gamma_{\rm dyn} < 1.0$), medium ($1.0 < \Gamma_{\rm dyn} < 2.0$) and  high ($\Gamma_{\rm dyn} > 2.0$) recent mass accretion rates, respectively. The stellar distribution shows a caustic feature coincident with the dark matter. This caustic is stronger and located at smaller radii for higher mass accretion rates.}
          \label{fig:dens_stack}
\end{figure*}

\subsection{Dark matter}
\label{sec:dm}
The logarithmic slope profiles of the dark matter density profiles for the $N=30$ C-EAGLE clusters are shown in Fig. \ref{fig:dm_slopes}. The clusters are ordered according to the recent mass accretion rate, $\Gamma_{\rm dyn}$, increasing from the top left panel to the bottom right panel. Here, we define mass accretion rate as:
\begin{equation}
\label{eq:gamma}
  \Gamma_{\rm dyn}(t) = \frac{\mathrm{log}\left[M_{\rm vir}(t)\right]-\mathrm{log}\left[M_{\rm vir}(t-t_{\rm dyn})\right]}{\mathrm{log}\left[a(t)\right]-\mathrm{log}\left[a(t-t_{\rm dyn})\right]}
\end{equation}
where $t_{\rm dyn}$ is the dynamical time, which corresponds to $z=0.5$ (or $a=0.667$) for redshift zero haloes \citep{diemer14, diemer17}.  Note, here we calculate the mass accretion rate using the virial mass, $M_{\rm vir}$, which is defined using the \cite{bryan98} formalism (with density contrast of $\Delta_{\rm c} \sim \! \! 102$ relative to the critical density at $z=0$). However, using $M_{\rm 200m}$ instead of $M_{\rm vir}$ makes a very small difference \citep{xhakaj20}.
For the majority of systems, a prominent `dip' is seen in the slope profiles, which previous works have labelled as the splashback radius \citep[e.g.][]{diemer14, more15}. In this work, we also define this radius of steepest slope as the splashback, and indicate these with the solid blue lines in Fig. \ref{fig:dm_slopes}. In agreement with previous work, this feature tends to become more pronounced at higher mass accretion rates. We note that CE-05 is currently undergoing a major merger, and hence the splashback feature, particularly in the dark matter distribution, is washed out. 

In some cases, we also identify a secondary caustic feature in the density profiles (shown with the dotted blue lines). These are located at smaller radii, and have shallower slopes than the splashback. \cite{deason20} labelled these features as `second caustics', and we adopt this terminology here.  Note, however, that these features do not necessarily relate to the classical definition of second caustic from spherical (or ellipsoidal) collapse models (see e.g. \citealt{adhikari14}), and could have multiple origins. As these features are much weaker than the splashback, we must caution against fitting to noise. To this end, we only consider second caustics that have slopes steeper than $-2.5$ and the difference between the local minimum and maximum is greater than 0.5 dex. Interestingly, the second caustics tend to be more common amongst haloes with low mass accretion rates, which is what the \cite{adhikari14} models predict. However, the numbers are too small to make a definitive statement. The second caustic features certainly deserve further scrutiny, and this will be a topic of future work. In this work, we focus on the splashback radii, and now turn our attention to the stellar distribution.

\subsection{Stars}

In Fig. \ref{fig:star_slopes}, we show the logarithmic slope profiles for the stellar material. Here, we consider all stars in the cluster, and do not try to distinguish between the brightest cluster galaxy, diffuse stellar material, or the stars bound in subhaloes. Any biases caused by massive substructures are mitigated by the angular median method used to calculate the density profiles. Note, however, that these profiles are not `pure' ICL, as we have not explicitly removed stars bound to subhaloes. There are a variety of different definitions of the ICL in the literature, which can lead to significant differences in the derived ICL properties \citep[see e.g.][]{rudick11, montes19}. Here, we consider all distant halo stars, and use the angular median method to minimise the effects of massive substructures and other non-diffuse structures. In practice, this approach is appealing as it can, potentially, be used in \textit{both} simulations and observations. Conversely, simply removing all stars that are bound to subhaloes is an approach that is not directly applicable to observations. Furthermore, the identification of bound subhaloes depends on the algorithm used, and the resolution of the simulation (see Section \ref{sec:icl} for further discussion).

The stellar profiles in Fig. \ref{fig:star_slopes} look similar to the dark matter profiles: a prominent dip is seen in almost all cases, and in some cases a second caustic-type feature is also apparent.  There are, however, some differences. Most notably the scales in Fig. \ref{fig:star_slopes} are different. While the outer caustics in the dark matter tend to have slopes of $\sim \! -4.5$, the stellar caustics are much steeper, with steepest slopes around $-6.7$. Note that this is not simply due to the entire stellar distribution having a steeper density profile (i.e. a vertical shift in the logarithmic slope profiles). In fact, the stars have similar slopes to the dark matter at smaller radii, and are only steeper by $\sim \! \! 0.5-1$ dex (see also \citealt{schaller15, montes18, pillepich18}). Thus, although the stellar profiles are generally a bit steeper, the caustics are also more prominent. We also note that the spread of steepest slopes is larger for the stars; the dark matter profiles typically have slopes of $-4.5 \pm 0.6$, while the stars have slopes of $-6.7 \pm 1.5$. Note, as the stellar profiles are typically steeper than the dark matter, our criteria for identifying second caustics is slightly different. Namely, we only consider second caustics that have slopes steeper than $-4.0$ and the difference between the local minimum and maximum is greater than 0.75 dex. Finally, it is worth remembering that the absolute values of these steepest slopes depends on how they are measured. In particular, the window size used in the Savitzky--Golay smoothing algorithm can change the measured slope substantially. Thus, while the relative differences between the dark matter and stars are robust, the absolute values of the steepest slope must be taken with a grain of salt.

In Fig. \ref{fig:dens_stack}, we consider the stacked density profiles of both the dark matter (solid black lines) and stars (long-dashed red lines). Here, all of the systems are stacked in the left-hand panel, and the remaining panels show subsets of low (middle left, $\Gamma_{\rm dyn} < 1.0$), medium (middle right, $1<\Gamma_{\rm dyn} < 2$), and high (right, $\Gamma_{\rm dyn} > 2$) mass accretion rates. In each panel, we give the estimated caustic radius and associated uncertainty. Here, we use a bootstrap method (without replacement) to estimate the uncertainty in the caustic for the stacked profiles. Two things are immediately obvious from this figure: (1) The location of the `splashback' in dark matter coincides with the steepest slope of the stars, i.e., a `stellar splashback', and (2) the location and strength of this splashback radius, in both dark matter and stars varies with mass accretion rate: the caustic is stronger and located at smaller radii for higher mass accretion rates. We investigate these two key points in the following subsection.

\subsection{The stellar splashback}
\begin{figure}
 \centering
        \includegraphics[width=\linewidth,angle=0]{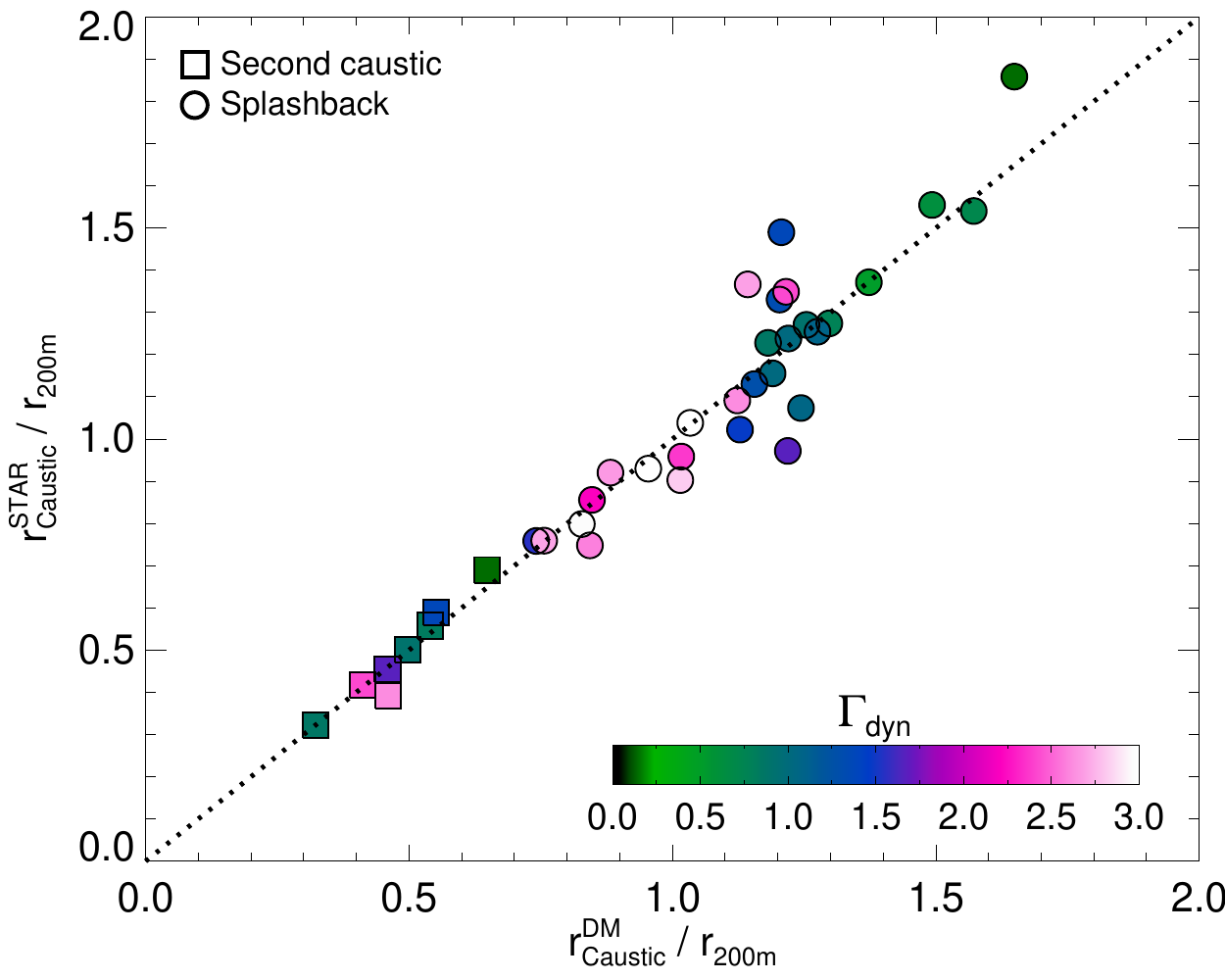}
         \caption[]{The location of the caustics in the stellar distribution against the caustic in the dark matter. The filled circles indicate the most prominent, outermost caustic, otherwise known as the splashback radius for dark matter. The square symbols show the (small number) of cases where a second caustic is identified in both the dark matter and the stars. The symbols are colour coded according to the recent mass accretion rate ($\Gamma_{\rm dyn}$, see Section \ref{sec:dm}). The dotted line shows the one-to-one relation: the stellar caustics are located at almost the same radius as the dark matter. As seen in previous work, the caustics tend to be located at smaller radii when the recent mass accretion rate is higher.}
          \label{fig:caustics}
\end{figure}

The apparent coincidence between the dark mater and stellar splashback radius seen in the stacked profiles is compelling. However, in order to determine whether or not these two radii are really related, we need to compare each individual halo. This is shown in Fig. \ref{fig:caustics}, where the stellar caustics are shown against the dark matter caustics. Here, the filled circle symbols  indicate the splashback radii, and, for completeness, the filled squares show the second caustics. Note that we only show second caustics when one is robustly identified in both the dark matter and stars; this occurs in $N=8$ haloes (27\%). In contrast, splashback radii in both stars and dark matter are found for all but one halo (CE-05 being the exception, which is currently undergoing a major merger). Remarkably, the stellar and dark matter caustics follow a tight one-to-one relation (with rms scatter $\Delta \left(R/r_{200\rm m}\right) = 0.11$, and Pearson/Spearman correlation coefficient 0.92/0.88). The points in Fig. \ref{fig:caustics} are colour coded according to the mass accretion rate. Here, we can see the trend alluded to in the previous figures: the splashback is located at smaller radii for higher mass accretion rates. We look at this more explicitly below.

\begin{figure}
 \centering
        \includegraphics[width=\linewidth,angle=0]{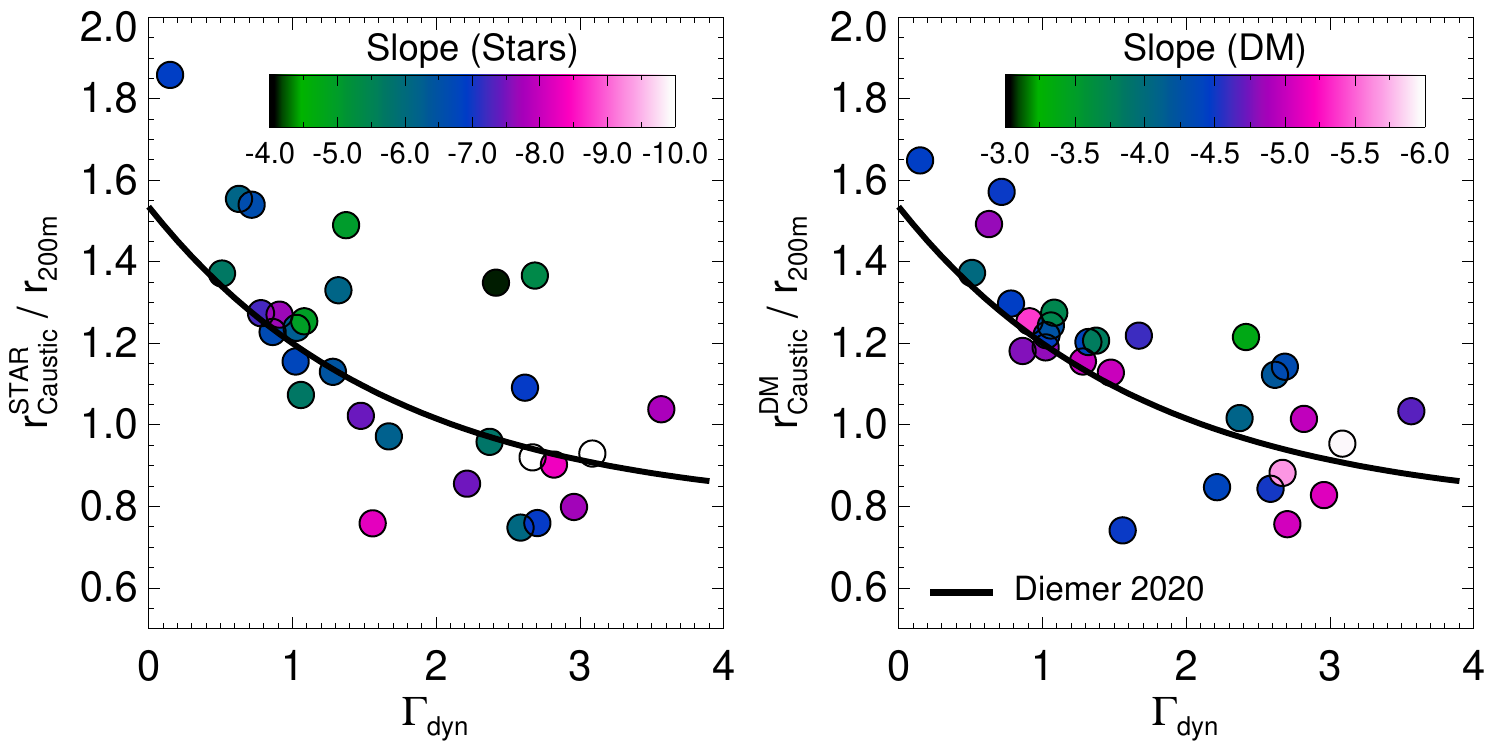}
         \caption[]{The location of the splashback caustics in the stellar distribution (left-hand panel) and the dark matter (right-hand panel) against the recent mass accretion rate. The points are coloured according to the density slope at the caustic. Haloes with higher mass accretion rates tend to have smaller splashback radii and steeper slopes. The solid black line shows the relation between splashback radius and mass accretion rate given by \cite{diemer20} for the median halo mass of the C-EAGLE sample.}
          \label{fig:caustics_gamma}
\end{figure}

Apart from being a more physically meaningful halo boundary, one of the most compelling reasons to probe the splashback radius in galaxy haloes is due to its strong link with mass accretion rate. Indeed, measurements of this radius can be used to classify galaxies by mass accretion rate, and can thus be used to probe aspects of halo formation like assembly bias (see e.g. \citealt{more16, busch17}). In Fig. \ref{fig:caustics_gamma}, we show the location of the stellar (left) and dark matter (right) splashback radii as a function of $\Gamma_{\rm dyn}$. The points are coloured according to the steepest slope at the caustic. The solid black line shows the predicted relation from \cite{diemer20}, assuming the median halo mass of the C-EAGLE sample ($\mathrm{log}_{10} M_{200 \rm m} /\mathrm{M}_\odot= 14.8$). Our results are in good agreement with the \cite{diemer20} predictions, and the stellar and dark matter caustics have $0.26$ and $0.14$ dex scatter in $r_{\rm Caustic}/r_{\rm 200m}$ about fixed $\Gamma_{\rm dyn}$, respectively. Perhaps most remarkable, however, is that the stellar caustics follow the \cite{diemer20} trend (albeit with slightly larger scatter than the dark matter). Indeed, these results suggest that not only can detection of an outer caustic in the stars be used to define the physical boundary of the halo, the stellar splashback radius can also be used to measure the mass accretion rate when $r_{200\rm m}$ is known. 

So far, we have only considered 3D density distributions. In reality, these are measured in projection, and thus to make connections with current and future observations we explore the projected density profiles in Section \ref{sec:icl}.

\begin{figure*}
 \centering
        \includegraphics[width=\linewidth,angle=0]{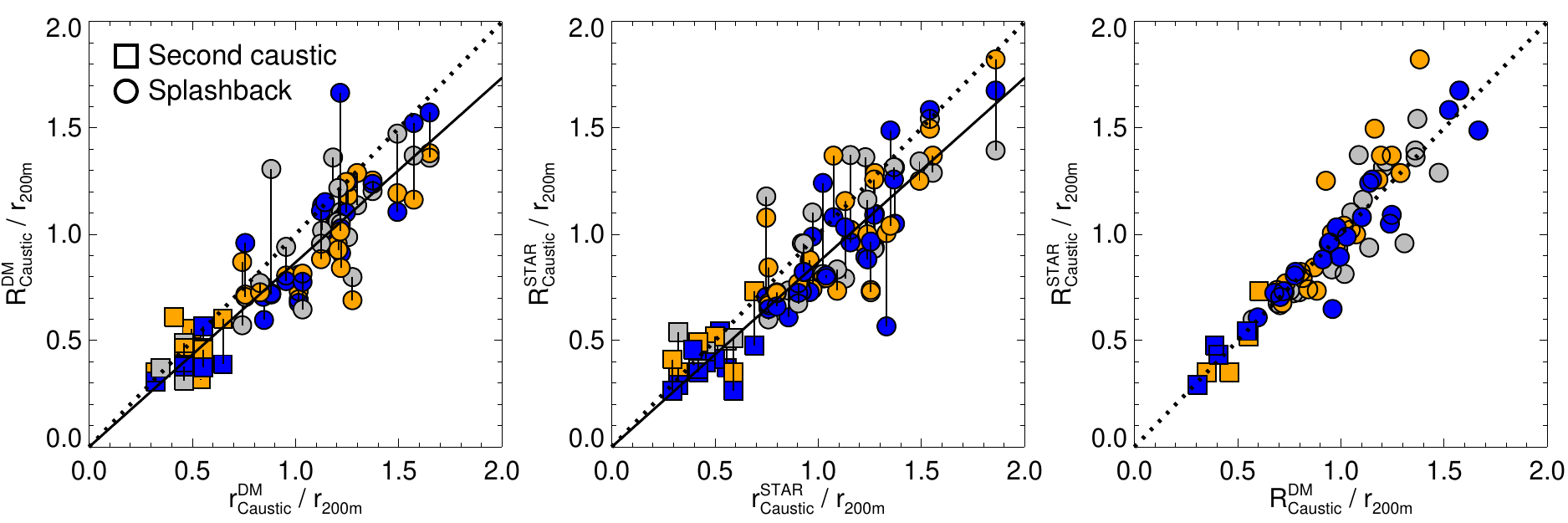}
        \caption[]{The location of the caustics for individual haloes in 2D projection ($R_{\rm Caustic}$). The left-hand and middle panels show the dark matter and stellar caustics in 2D projection versus 3D ($r_{\rm Caustic}$), respectively. The different coloured points show different projections (i.e. projections along $x$, $y$, or $z$ in the simulation box are shown with blue, orange, and grey points, respectively). The caustics in projection are typically $\sim \! \! 0.9$ times the 3D radius. The $R_{\rm Caustic} = 0.9 r_{\rm Caustic}$ relation is shown with the solid line, and the one-to-one relation is shown with the dotted line. The right-hand panel shows the 2D caustics for the stars versus the dark matter. As for the 3D case, these closely follow a one-to-one relation.}
          \label{fig:caustics_proj}
\end{figure*}

\subsection{Comparison with Milky Way-mass scales}
Before turning to the observational consequences of these theoretical results, it is worth discussing \textit{why} we see these stellar splashback features in the simulations. The dark matter density profile, and the associated splashback radius, have been studied extensively in previous works. However, the corresponding stellar profiles have received much less attention. This is perhaps unsurprising: the hydrodynamical simulations required to form stars are far more expensive than dark-matter-only simulations, and, perhaps more importantly, include uncertain subgrid galaxy formation prescriptions. \cite{deason20} studied the edges of stellar haloes using high-resolution simulations of Milky Way-mass galaxies. However, the contrast with the results for cluster-mass scales is striking! In particular, \cite{deason20} found that the stars did not generally reach out to the splashback radius of the dark matter, and, in fact, the edge of the Galactic-sized stellar haloes more often coincide with the second caustic of the dark matter. We suggest that this difference is mainly owing to three mass-dependent effects: (1) the stellar mass--halo mass relation, (2) the importance of smooth accretion, and (3) the formation age or concentration of the host halo. 

First, the stellar mass--halo mass relation is non-linear and varies as a function of halo mass \citep{moster10, behroozi13}. Milky Way-mass haloes accrete most of their mass from small subhaloes, which themselves have high dark matter fractions, and, in some cases, no stars at all. On the other hand, the diffuse light on cluster-mass scales is dominated by the remains of massive galaxies ($\sim \! \! 10^{10}-10^{12} \mathrm{M}_\odot$), which form stars efficiently \citep[see e.g.][]{conroy07, purcell07, puchwein10}. This leads to Galactic stellar haloes being dominated by a small number of progenitors \citep[see e.g.][]{deason16}, while a larger number of progenitors contribute to the ICL. Thus, the stellar mass--halo mass relation can partly explain why the stellar density profiles of cluster-mass haloes are more strongly related to the underlying dark matter distribution \citep[e.g.][]{montes18, pillepich18}. Second, the importance of smooth accretion, in the form of dark matter particles not bound to any halo, varies as a function of halo mass; smooth accretion is dominant on Milky Way-mass scales, but mergers dominate the mass growth in clusters \citep[e.g.][]{fakhouri10, genel10, wang11}. Moreover, the fraction of mass in substructures is much larger in clusters than galactic haloes \citep[e.g.][]{gao11}. Thus, there are many more massive objects losing stars in a cluster than there would be in a typical Milky Way-mass halo. Finally, the cluster-mass haloes tend to form later, and have lower concentration than Milky Way-mass haloes. The luminous satellites that are accreted more recently tend to deposit stars at larger radii \citep[see e.g.][]{cooper15}, and thus the stripped material can reach out to the splashback radius. Note, here we have discussed the main factors that we believe determine the location of the stellar edges on different mass scales. However, there are many other mass-dependent effects that could be important. For example, the relevance of pre-processed satellite galaxies \citep[e.g.][]{bahe19}, the (stellar and dark) density profiles of the disrupting satellites \citep[e.g.][]{penarrubia08,watson12}, and the survival times of satellite galaxies \citep[e.g][]{bahe19}.

We end this interlude by noting that the results found here for cluster-mass haloes and the previous \cite{deason20} work on Milky Way-mass scales, span a significant mass range, but further work is warranted to fill in the remaining `mass-gap'. For example, does the stellar edge smoothly change with radius between the second caustic of the dark matter and the splashback radius, or is there a sudden transition at a particular mass scale? This, and other related questions encourage a separate, more extensive, study of stellar halo edges across a range of halo masses.

\section{Projected profiles}

\label{sec:icl}
\begin{figure*}
 \centering
        \includegraphics[width=\linewidth,angle=0]{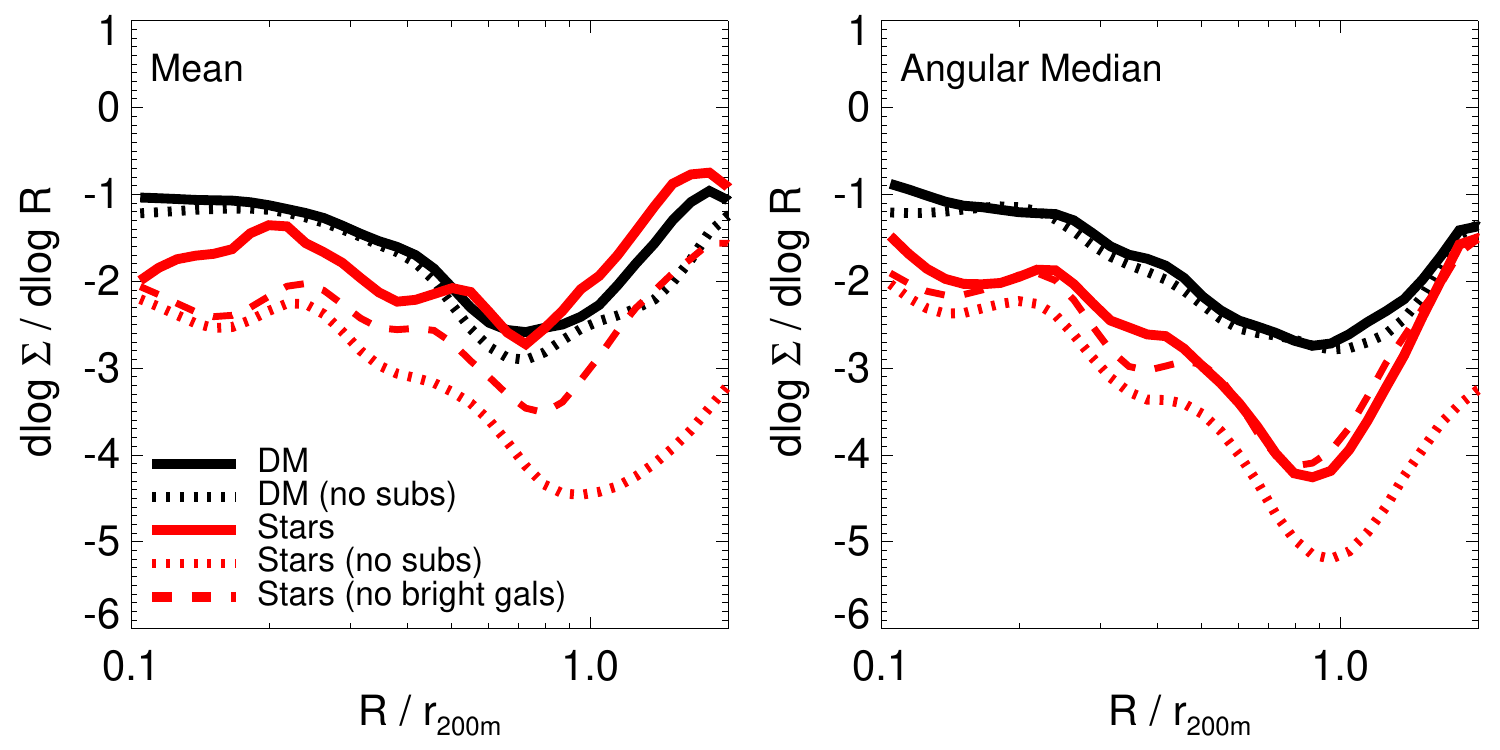}
         \caption[]{The stacked logarithmic slope profiles of the projected dark matter (black) and stellar (red) density profiles of the C-EAGLE clusters. Here, the projected density profiles of the clusters are computed using two different methods: the mean density (left), and the angular median density (right). The dotted lines show the profiles when all substructures are removed, and the dashed red lines show the stellar profiles when bright satellite galaxies ($M_{\rm Star} > 10^9M_\odot$) are excluded. The caustics for both dark matter and stars are weaker when the mean profile is used. In addition, the influence of substructures is more pronounced in the stellar profiles, which significantly contribute to the star light at large radii.}
          \label{fig:dens_stack_proj}
\end{figure*}

In this section, we investigate the projected density profiles, which are more relevant for observational studies of clusters.
Projected density profiles are constructed in a similar way to the 3D profiles. We use the same radial bin size and smoothing, and, by default, compute the projected density in each radial shell using the angular median method. In the angular median method used above, each radial bin in the 3D density profile is split into 50 evenly spaced solid angles, and the median value is computed. For the 2D projected profiles, the same number of angular bins are used, but these bins are angles instead of solid angles. As we are considering particles within a $4r_{\rm 200m}$ spherical aperture, there can be projection effects at larger radii (where we are artificially running out of particles). However, we find that these effects are minimal within $2r_{200\rm m}$, within which the outer caustics are typically located.

In Fig. \ref{fig:caustics_proj}, we show the dark matter (left-hand panel) and stellar (right-hand panel) caustics for individual haloes in 2D projection versus 3D. The filled circles show the splashback radii (or outer caustic) and the filled squares show, where applicable, the second caustics. Three different coloured symbols are used to indicate projections computed along different axes (in this case, along $x$, $y$, or $z$ in the simulation box). The results for the different projections of a halo are connected with a solid vertical line. In some cases, these differences can be substantial. The dotted lines in the panels show the one-to-one relation, but we also show a solid line which best describes the relation between the projected and 3D quantities where, $R_{\rm caustic} \sim \! 0.9 r_{\rm caustic}$. Finally, in the right-hand panel, we show the projected caustics for stars versus the dark matter. Like the 3D cases, these caustics line up on the one-to-one line and are directly related.

\begin{figure*}
 \centering
        \includegraphics[width=\linewidth,angle=0]{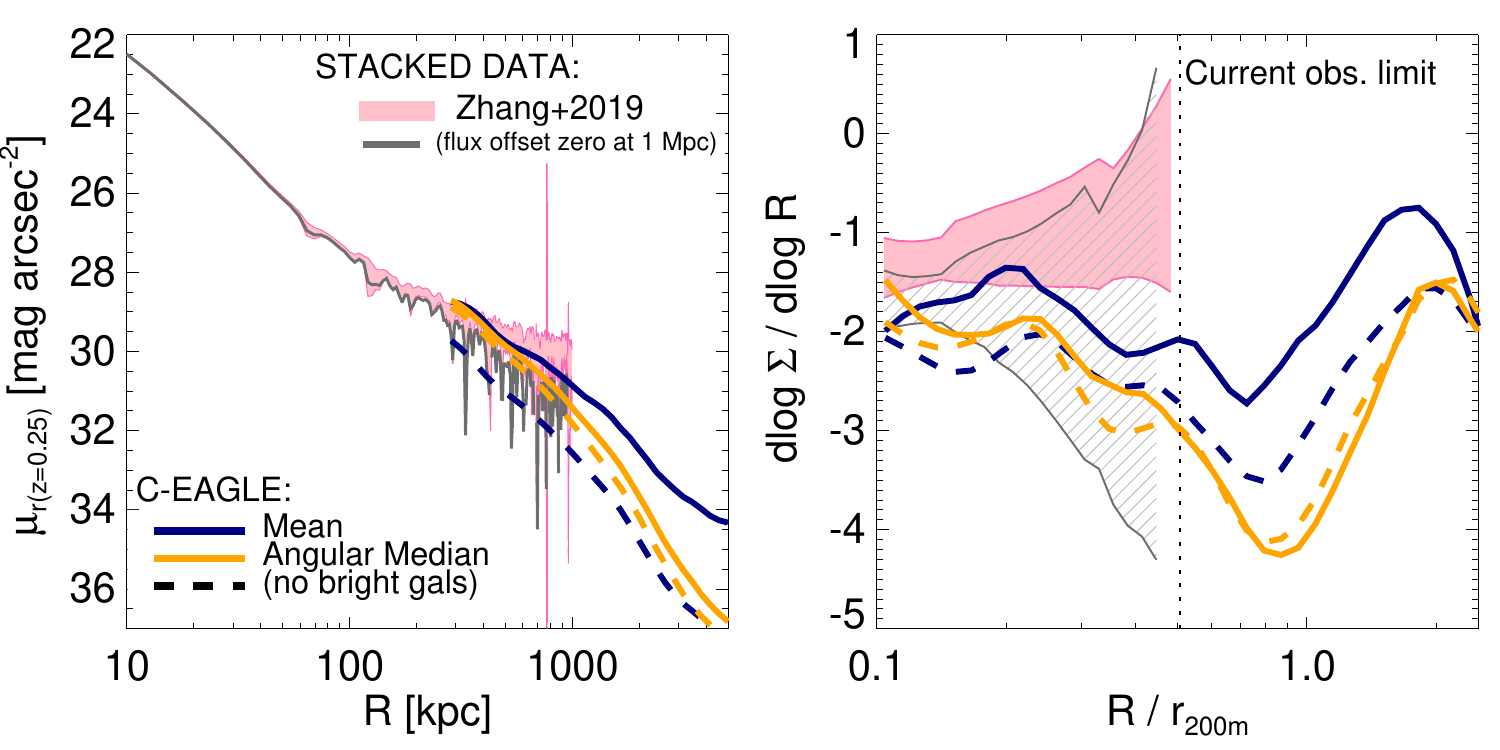}
         \caption[]{\textit{Left:} the stacked surface brightness profile of $z=0.25$ DES clusters from \cite{zhang19}. The fiducial \cite{zhang19} profiles are shown with the pink shaded region, and the dark grey line is the same data with a different zero flux offset. This latter flux offset was chosen to be similar to the SDSS \cite{zibetti05} stacks. The projected stellar mass density profiles of the C-EAGLE clusters have been converted into surface brightness assuming a constant mass-to-light ratio of $M/L=5$, and then scaled to the same (average) halo mass as the DES sample. We have also included the $(1+z)^4$ dimming factor applicable for $z=0.25$. We show the mean and the angular median stacked profiles with the dark blue and orange lines, respectively. The corresponding profiles when bright galaxies are removed are shown with the dashed lines. \textit{Right:} the logarithmic slope profiles of the projected stellar density profiles for the C-EAGLE clusters and the data. Identifying the caustic feature in the light distribution requires probing to larger projected distances, and thus fainter surface brightness limits ($\mu \sim  \!  34$ mag arcsec$^{-2}$ at $z=0.25$). In addition, the caustic will only be detected if stacking methods take into account the influence of outliers, for example by using an angular median method.}
          \label{fig:surf_brightness}
\end{figure*}

Up to now, we have computed density profiles for individual clusters using an angular median method \citep{mansfield17}. In Fig. \ref{fig:dens_stack_proj}, we show the stacked profiles when two different methods are used to compute the \textit{individual} profiles: (1) the mean in each (projected) radial shell, (2) the angular median density in each (projected) radial shell; our fiducial method.  Unsurprisingly, the caustics for both the dark matter and the stars are weaker when the mean density profile is used. Indeed, this is one of the reasons the angular median method was proposed by \cite{mansfield17}, as the mean values in radial shells are more affected by outliers. However, it is also apparent that the difference between the mean and angular median profiles is much more relevant for the stellar material. The stacked caustic is hardly identifiable with the mean profile, but is very prominent when the angular median profile is used. The reason for this becomes clear when we consider profiles with bright galaxies explicitly removed (shown with the red dashed lines). Here, we have excluded all star particles bound to a subhalo with $M_{\rm Star} > 10^9M_\odot$. This is approximately the stellar mass limit for which cluster member galaxies can be masked in observations \citep[e.g.][]{zibetti05, zhang19}.  When the bright galaxies are excluded the splashback feature is discernible, even when the mean method is used. However, the angular median profile is largely unchanged, and is still more prominent than the mean profile. This is because there are additional structures, such as streams, plumes and clouds, that can affect the derived density profile. Thus, using a technique such as the angular median method is \textit{essential} in order to detect the edge of the stellar material. Note that stacking a significant amount of systems could help alleviate this problem (we only have 30 haloes to stack with C-EAGLE), but the method used to measure the density profiles will still affect how strong the derived signal, if any, is.

The dotted lines in Fig. \ref{fig:dens_stack_proj} show the stacked profiles when \textit{all subhaloes} identified by the \textsc{subfind} algorithm \citep{springel01} are omitted. The outer profiles are steeper when subhaloes are removed (see e.g. \citealt{fielder20}), however, this makes a much larger difference to the stellar density profiles. This is because the fraction of stars bound in subhaloes at large radii is much greater than the fraction of dark matter \citep[e.g.][]{gao11,pillepich18}. As we discussed earlier, the process of removing stars bound to satellite galaxies cannot be directly replicated in observations (and, importantly, the identification of subhaloes in the simulations is not perfect, e.g. \citealt{canas19}). Relatively bright galaxies can be identified, but the contribution of fainter satellites is either ignored, or roughly estimated by making assumptions about the satellite galaxy luminosity function and their radial distribution \citep[e.g.][]{zibetti05,zhang19,sampaio20}. Fortunately, when the angular median method is used, the stellar splashback feature is present with or without the contribution of satellite galaxies. However, in order to robustly compare the location of this splashback feature with observations (see the following section), it will be desirable to perform mock observations where the cluster light profiles are computed in the same way as the data.

\subsection{Observations}
The results of the previous sections predict a well-defined `edge' at the outskirts of the stellar distribution of cluster-mass haloes. Recent work has shown that the ICL closely follows the total matter distribution \citep{montes19, asensio20}. In agreement with these results, we have now shown that the stars at the very outskirts of the cluster have a splashback radius, or radius of steepest slope, at the same location as the dark matter. This prediction begs the question: can this edge be observed in the ICL?
 
In Fig. \ref{fig:surf_brightness} we show the surface brightness profile for the stacked sample of $z=0.25$ clusters from \cite{zhang19}. This profile was derived from $N=300$ galaxy clusters from the Dark Energy Survey (DES) Year 1 data, with median halo mass of $M_{200 \rm m} = 2.5 \times 10^{14}\mathrm{M}_\odot$. The shaded pink region shows their fiducial profile, while the gray line indicates the profile when the flux is defined to be zero at 1 Mpc (cf. zero flux at 1.8 Mpc for the fiducial profile). This latter profile was made to directly compare with the \cite{zibetti05} results from a stack of SDSS clusters.
The dark blue and orange lines show the (median) surface brightness profile for the C-EAGLE clusters using the mean, or angular median method to compute the density profiles. The dashed lines show the profiles when bright galaxies have been explicitly removed. The projected stellar mass density profiles of the C-EAGLE clusters have been converted into surface brightness assuming a constant mass-to-light ratio of $M/L=5 \mathrm{M}_\odot/\mathrm{L}_\odot$ (as in \citealt{asensio20}). In addition, we have scaled the derived intensity profiles to the median halo mass of the DES sample surface mass density, assuming the stellar mass of the ICL scales with halo mass (see e.g. \citealt{pillepich18}). Finally, we have included the $(1+z)^4$ dimming factor to take into account the $z=0.25$ redshift of the observed clusters. Note that the \textit{absolute} surface brightness level of the C-EAGLE clusters should be taken with a pinch of salt, as this depends on the agreement with the observed stellar masses and the exact way in which the ICL is defined. Nonetheless our simple conversion of projected surface density to surface brightness shows reasonable agreement with the observed data, and gives a good indication of the surface brightness levels needed to probe to the cluster edges.

In the right-hand panel of Fig. \ref{fig:surf_brightness}, we show the logarithmic slope profiles of the projected stellar density profiles. The lines from C-EAGLE are taken from Fig. \ref{fig:dens_stack_proj}. We show the two profiles from \cite{zhang19} using different flux offsets with the filled pink and line-filled gray regions, respectively. For the observed profile, we scale relative to the median $r_{200 \rm m}$ of the sample. The errors in the observed fluxes are taken into account by computing the slope profiles many ($N=10^4$) times and scattering the values according to the error distributions. Here, we ignore bins where the errors are particularly high (greater than 10\%). In addition, we rebin the observed data to have $N=21$ logarithmic radial bins between $-1.0 < \mathrm{log} \left(R/r_{200 \rm m}\right) < -0.3$.

It is clear from Fig. \ref{fig:surf_brightness} that, regardless of the flux offset, the DES data are consistent with a constant slope. However, even though the data reach to an impressive 1 Mpc, in order to probe the predicted stellar splashback, the data would need to go out to at least 2 Mpc --- or surface brightness levels of $\mu \sim \! 32-36$ mag arcsec$^{-2}$. Although this appears to be unfeasible with current observations, this is certainly achievable with future observations. The Vera C. Rubin Observatory Legacy Survey of Space and Time (LSST, \citealt{lsst}) will survey an area that is $> 10$ times larger than the DES Y1 footprint, and is predicted to reach a depth of at least 2 mag deeper. With the increase in both area and depth, and hence an increase in ICL flux collection by a factor of $\sim \! 100$, LSST is capable of reaching these extreme projected distances and surface brightness levels.
In the shorter term, the increased area and depth from the final DES data release may also provide some constraints on the stellar splashback; i.e., the DES data may be able to rule out such a feature if the profile does not show the predicted sharp drop. The upcoming spaced-based Euclid mission \citep{euclid} will probe thousands of cluster systems, and can also be used to create deep, stacked ICL profiles. The Euclid-wide survey will cover a similar area to LSST (Euclid-wide: $15 000$ deg$^2$, LSST: $18 000$ deg$^2$), but it is significantly shallower (by 2-3 mag). However, like the final DES release, this survey can still provide important constraints on the splashback radii of clusters. It is also worth considering the methods used to create the stacked ICL profiles. For example, while care is often taken to remove bright galaxies from the profiles, further improvement could be made by applying an angular median method on each individual cluster, as used in this work. 

In addition to stacked profiles, there is also scope to probe these extreme outskirts for individual clusters. The Beyond Ultra-deep Frontier Fields and Legacy Observations (BUFFALO) Hubble Space Telescope (\textit{HST}) treasury program \citep{buffalo} can probe the ICL of individual clusters out to $\sim \! \! 1$ Mpc, and an extension of this program could push to even larger distances. For individual clusters, care must be taken to avoid confusion between the stellar splashback with the second caustic (this secondary feature is normally washed out in stacked profiles). However, these secondary features tend to be less prominent amongst haloes with very high (recent) mass accretion rates, and the BUFFALO clusters are particularly active systems. Finally, the Nancy Grace Roman Space Telescope \citep{wfirst} is the obvious successor to \textit{HST} for probing the ICL of individual systems. The BUFFALO survey is already capable of probing out to similar projected distances for individual systems as stacked profiles from DES, and the $100$ times larger field-of-view of Roman will enable programs such as BUFFALO to probe out to the (predicted) splashback radius.

\section{Conclusions}
\label{sec:conc}
We have used the C-EAGLE suite of simulations to explore the outskirts of dark matter and stars on cluster-mass scales. Density profiles of each C-EAGLE system are constructed using an angular median method, which limits the influence of substructure and other non-diffuse components. The radius of steepest slope is used as a proxy for the splashback radius, which corresponds to the first apocentre of recently infalling material. The outer caustics, or splashback radii, of both the dark matter and stellar components are compared, both for individual clusters and with stacks based on mass accretion rate. Our main conclusions are summarised as follows:

\begin{itemize}
\setlength\itemsep{1em}

\item The stellar density profiles of clusters have a well-defined edge, defined by the radius of steepest log-density slope, which coincides with the outer dark matter caustic, or the splashback radius. This radius is typically located at $r_{200 \rm m}$, in good agreement with previous work using dark-matter-only simulations.

\item The location of the stellar and dark matter splashback radius depends on the mass accretion rate: slowly accreting haloes tend to have an edge at a larger radius, and a shallower steepest slope. The stellar profiles have more prominent outer caustics than the dark matter. In some cases ($\sim \! \! 27\%$), a secondary caustic can be identified in the stellar and dark matter profiles: these likely correspond to the apocentre of material that has completed at least two pericentric passages, but the features are much weaker than the radius of steepest slope, and hence more difficult to detect.

\item The radius of steepest slope can also be identified in projection, where the 2D and 3D radii are related by $R_{\rm Caustic} \sim \! 0.9 r_{\rm caustic}$. The method used to identify the caustic is crucial, as massive substructures can significantly dilute the signal of the steepest slope. This is especially true for the stellar material: there is a higher fraction of stars than dark matter bound in subhaloes (see e.g. \citealt{pillepich18}).

\item Current observations of the ICL can reach out to $\sim \! \! 1$ Mpc, either for individual systems or from stacking many systems. Detecting the stellar splashback will require probing out a further 1 Mpc, to surface brightnesses of $\mu \sim \! \! 32-36$ mag arcsec$^{-2}$. However, this challenging feat will be achievable with upcoming facilities ideally suited to low surface brightness studies, such as the Vera C. Rubin Observatory \citep[see e.g.][]{brough20}, the Nancy Grace Roman Space Telescope, and Euclid.

\end{itemize}

The relation between the visible and dark matter is complex, mass-dependent and depends on galaxy formation physics, hierarchical structure formation and cosmology. The stellar haloes of galaxies and clusters offer a unique way to probe the dark matter, as these are mainly built from mergers. This work shows that measuring the `edge' of the ICL offers an alternative way to define the halo boundary and quantify the mass accretion rate of clusters. Moreover, by comparing the stellar splashback with independent measures of the splashback radius, e.g. from satellite galaxies or weak lensing, we can quantify the link between the stellar and dark material, and thus test the predictions of our galaxy formation models. Ultimately, learning how the outskirts of dark and stellar haloes change with mass and time will provide an invaluable way to critically examine, and inform, our state-of-the-art cosmological models of structure formation.

\section*{Acknowledgements}
We thank the referee of this paper, Benedikt Diemer, for providing insightful and constructive comments and suggestions.
AD also thanks Anthony Gonzalez and Richie Wang for providing useful comments that have improved this manuscript. Our gratitude is extended to all of the essential workers that support our livelihood, especially during the Coronavirus pandemic. Last, and certainly not least, AD thanks the staff at the Durham University Day Nursery who play a key role in enabling research like this to happen.

AD is supported by a Royal Society University Research Fellowship, and by the Science and Technology Facilities Council (STFC)(grant numbers ST/P000541/1, and ST/T000244/1). AD and AF also acknowledge support from the Leverhulme Trust. KAO acknowledges support by the European Research Council (ERC) through Advanced Investigator grant DMIDAS (GA 786910). MS is supported by the Netherlands Organisation
  for Scientific Research (NWO) through VENI grant 639.041.749. MJ is supported by the United Kingdom Research and Innovation (UKRI) Future Leaders Fellowship `Using Cosmic Beasts to uncover the Nature of Dark Matter' (grant number MR/S017216/1). YMB acknowledges funding from the EU Horizon 2020 research and innovation programme under Marie Skłodowska-Curie grant agreement 747645 (ClusterGal) and the Netherlands Organisation for Scientific Research (NWO) through VENI grant 639.041.751. CDV acknowledges the support of the Spanish Ministry of Science, Innovation and Universities (MCIU) through grants RYC-2015-18078 and PGC2018-094975-B-C22. This work used the DiRAC@Durham facility managed by the Institute for Computational Cosmology on behalf of the STFC DiRAC HPC Facility (\url{www.dirac.ac.uk}). The equipment was funded by BEIS capital funding via STFC capital grants ST/K00042X/1, ST/P002293/1, ST/R002371/1, and ST/S002502/1, Durham University and STFC operations grant ST/R000832/1. DiRAC is part of the National e-Infrastructure. The C-EAGLE simulations were in part performed on the German federal maximum performance computer `HazelHen' at the maximum performance computing centre Stuttgart (HLRS), under project GCS-HYDA / ID 44067 financed through the large-scale project Hydrangea' of the Gauss Center for Supercomputing. Further simulations were performed at the Max Planck Computing and Data Facility in Garching, Germany.

\section*{Data Availability Statement}
The data presented in the figures are available upon request from the corresponding author. The raw simulation data can be requested from the C-EAGLE team \citep{bahe17, barnes17}.

\bibliographystyle{mnras}
\bibliography{mybib}

\begin{thebibliography}{}
\makeatletter
\relax
\def\mn@urlcharsother{\let\do\@makeother \do\$\do\&\do\#\do\^\do\_\do\%\do\~}
\def\mn@doi{\begingroup\mn@urlcharsother \@ifnextchar [ {\mn@doi@}
  {\mn@doi@[]}}
\def\mn@doi@[#1]#2{\def\@tempa{#1}\ifx\@tempa\@empty \href
  {http://dx.doi.org/#2} {doi:#2}\else \href {http://dx.doi.org/#2} {#1}\fi
  \endgroup}
\def\mn@eprint#1#2{\mn@eprint@#1:#2::\@nil}
\def\mn@eprint@arXiv#1{\href {http://arxiv.org/abs/#1} {{\tt arXiv:#1}}}
\def\mn@eprint@dblp#1{\href {http://dblp.uni-trier.de/rec/bibtex/#1.xml}
  {dblp:#1}}
\def\mn@eprint@#1:#2:#3:#4\@nil{\def\@tempa {#1}\def\@tempb {#2}\def\@tempc
  {#3}\ifx \@tempc \@empty \let \@tempc \@tempb \let \@tempb \@tempa \fi \ifx
  \@tempb \@empty \def\@tempb {arXiv}\fi \@ifundefined
  {mn@eprint@\@tempb}{\@tempb:\@tempc}{\expandafter \expandafter \csname
  mn@eprint@\@tempb\endcsname \expandafter{\@tempc}}}

\bibitem[\protect\citeauthoryear{{Adhikari}, {Dalal}  \&
  {Chamberlain}}{{Adhikari} et~al.}{2014}]{adhikari14}
{Adhikari} S.,  {Dalal} N.,   {Chamberlain} R.~T.,  2014, \mn@doi [\jcap]
  {10.1088/1475-7516/2014/11/019}, \href
  {https://ui.adsabs.harvard.edu/abs/2014JCAP...11..019A} {2014, 019}

\bibitem[\protect\citeauthoryear{{Adhikari}, {Dalal}  \& {Clampitt}}{{Adhikari}
  et~al.}{2016}]{adhikari16}
{Adhikari} S.,  {Dalal} N.,   {Clampitt} J.,  2016, \mn@doi [\jcap]
  {10.1088/1475-7516/2016/07/022}, \href
  {https://ui.adsabs.harvard.edu/abs/2016JCAP...07..022A} {2016, 022}

\bibitem[\protect\citeauthoryear{{Adhikari} et~al.,}{{Adhikari}
  et~al.}{2020}]{adhikari20}
{Adhikari} S.,  et~al., 2020, arXiv e-prints, \href
  {https://ui.adsabs.harvard.edu/abs/2020arXiv200811663A} {p. arXiv:2008.11663}

\bibitem[\protect\citeauthoryear{{Alonso Asensio}, {Dalla Vecchia}, {Bah{\'e}},
  {Barnes}  \& {Kay}}{{Alonso Asensio} et~al.}{2020}]{asensio20}
{Alonso Asensio} I.,  {Dalla Vecchia} C.,  {Bah{\'e}} Y.~M.,  {Barnes} D.~J.,
  {Kay} S.~T.,  2020, \mn@doi [\mnras] {10.1093/mnras/staa861}, \href
  {https://ui.adsabs.harvard.edu/abs/2020MNRAS.494.1859A} {494, 1859}

\bibitem[\protect\citeauthoryear{{Bah{\'e}} et~al.,}{{Bah{\'e}}
  et~al.}{2017}]{bahe17}
{Bah{\'e}} Y.~M.,  et~al., 2017, \mn@doi [\mnras] {10.1093/mnras/stx1403},
  \href {https://ui.adsabs.harvard.edu/abs/2017MNRAS.470.4186B} {470, 4186}

\bibitem[\protect\citeauthoryear{{Bah{\'e}} et~al.,}{{Bah{\'e}}
  et~al.}{2019}]{bahe19}
{Bah{\'e}} Y.~M.,  et~al., 2019, \mn@doi [\mnras] {10.1093/mnras/stz361}, \href
  {https://ui.adsabs.harvard.edu/abs/2019MNRAS.485.2287B} {485, 2287}

\bibitem[\protect\citeauthoryear{{Banerjee}, {Adhikari}, {Dalal}, {More}  \&
  {Kravtsov}}{{Banerjee} et~al.}{2020}]{banerjee20}
{Banerjee} A.,  {Adhikari} S.,  {Dalal} N.,  {More} S.,   {Kravtsov} A.,  2020,
  \mn@doi [\jcap] {10.1088/1475-7516/2020/02/024}, \href
  {https://ui.adsabs.harvard.edu/abs/2020JCAP...02..024B} {2020, 024}

\bibitem[\protect\citeauthoryear{{Barnes}, {Kay}, {Henson}, {McCarthy},
  {Schaye}  \& {Jenkins}}{{Barnes} et~al.}{2017a}]{barnes17b}
{Barnes} D.~J.,  {Kay} S.~T.,  {Henson} M.~A.,  {McCarthy} I.~G.,  {Schaye} J.,
    {Jenkins} A.,  2017a, \mn@doi [\mnras] {10.1093/mnras/stw2722}, \href
  {https://ui.adsabs.harvard.edu/abs/2017MNRAS.465..213B} {465, 213}

\bibitem[\protect\citeauthoryear{{Barnes} et~al.,}{{Barnes}
  et~al.}{2017b}]{barnes17}
{Barnes} D.~J.,  et~al., 2017b, \mn@doi [\mnras] {10.1093/mnras/stx1647}, \href
  {https://ui.adsabs.harvard.edu/abs/2017MNRAS.471.1088B} {471, 1088}

\bibitem[\protect\citeauthoryear{{Baxter} et~al.,}{{Baxter}
  et~al.}{2017}]{baxter17}
{Baxter} E.,  et~al., 2017, \mn@doi [\apj] {10.3847/1538-4357/aa6ff0}, \href
  {https://ui.adsabs.harvard.edu/abs/2017ApJ...841...18B} {841, 18}

\bibitem[\protect\citeauthoryear{{Behroozi}, {Wechsler}  \&
  {Conroy}}{{Behroozi} et~al.}{2013}]{behroozi13}
{Behroozi} P.~S.,  {Wechsler} R.~H.,   {Conroy} C.,  2013, \mn@doi [\apj]
  {10.1088/0004-637X/770/1/57}, \href
  {https://ui.adsabs.harvard.edu/abs/2013ApJ...770...57B} {770, 57}

\bibitem[\protect\citeauthoryear{{Brough} et~al.,}{{Brough}
  et~al.}{2020}]{brough20}
{Brough} S.,  et~al., 2020, arXiv e-prints, \href
  {https://ui.adsabs.harvard.edu/abs/2020arXiv200111067B} {p. arXiv:2001.11067}

\bibitem[\protect\citeauthoryear{{Bryan} \& {Norman}}{{Bryan} \&
  {Norman}}{1998}]{bryan98}
{Bryan} G.~L.,  {Norman} M.~L.,  1998, \mn@doi [\apj] {10.1086/305262}, \href
  {https://ui.adsabs.harvard.edu/abs/1998ApJ...495...80B} {495, 80}

\bibitem[\protect\citeauthoryear{{Budzynski}, {Koposov}, {McCarthy}, {McGee}
  \& {Belokurov}}{{Budzynski} et~al.}{2012}]{budzynski12}
{Budzynski} J.~M.,  {Koposov} S.~E.,  {McCarthy} I.~G.,  {McGee} S.~L.,
  {Belokurov} V.,  2012, \mn@doi [\mnras] {10.1111/j.1365-2966.2012.20663.x},
  \href {https://ui.adsabs.harvard.edu/abs/2012MNRAS.423..104B} {423, 104}

\bibitem[\protect\citeauthoryear{{Bullock} \& {Johnston}}{{Bullock} \&
  {Johnston}}{2005}]{bullock05}
{Bullock} J.~S.,  {Johnston} K.~V.,  2005, \mn@doi [\apj] {10.1086/497422},
  \href {https://ui.adsabs.harvard.edu/abs/2005ApJ...635..931B} {635, 931}

\bibitem[\protect\citeauthoryear{{Busch} \& {White}}{{Busch} \&
  {White}}{2017}]{busch17}
{Busch} P.,  {White} S. D.~M.,  2017, \mn@doi [\mnras] {10.1093/mnras/stx1584},
  \href {https://ui.adsabs.harvard.edu/abs/2017MNRAS.470.4767B} {470, 4767}

\bibitem[\protect\citeauthoryear{{Ca{\~n}as}, {Elahi}, {Welker}, {del P Lagos},
  {Power}, {Dubois}  \& {Pichon}}{{Ca{\~n}as} et~al.}{2019}]{canas19}
{Ca{\~n}as} R.,  {Elahi} P.~J.,  {Welker} C.,  {del P Lagos} C.,  {Power} C.,
  {Dubois} Y.,   {Pichon} C.,  2019, \mn@doi [\mnras] {10.1093/mnras/sty2725},
  \href {https://ui.adsabs.harvard.edu/abs/2019MNRAS.482.2039C} {482, 2039}

\bibitem[\protect\citeauthoryear{{Chang} et~al.,}{{Chang}
  et~al.}{2018}]{chang18}
{Chang} C.,  et~al., 2018, \mn@doi [\apj] {10.3847/1538-4357/aad5e7}, \href
  {https://ui.adsabs.harvard.edu/abs/2018ApJ...864...83C} {864, 83}

\bibitem[\protect\citeauthoryear{{Conroy}, {Wechsler}  \& {Kravtsov}}{{Conroy}
  et~al.}{2007}]{conroy07}
{Conroy} C.,  {Wechsler} R.~H.,   {Kravtsov} A.~V.,  2007, \mn@doi [\apj]
  {10.1086/521425}, \href
  {https://ui.adsabs.harvard.edu/abs/2007ApJ...668..826C} {668, 826}

\bibitem[\protect\citeauthoryear{{Contigiani}, {Hoekstra}  \&
  {Bah{\'e}}}{{Contigiani} et~al.}{2019}]{contigiani19}
{Contigiani} O.,  {Hoekstra} H.,   {Bah{\'e}} Y.~M.,  2019, \mn@doi [\mnras]
  {10.1093/mnras/stz404}, \href
  {https://ui.adsabs.harvard.edu/abs/2019MNRAS.485..408C} {485, 408}

\bibitem[\protect\citeauthoryear{{Contini}, {De Lucia}, {Villalobos}  \&
  {Borgani}}{{Contini} et~al.}{2014}]{contini14}
{Contini} E.,  {De Lucia} G.,  {Villalobos} {\'A}.,   {Borgani} S.,  2014,
  \mn@doi [\mnras] {10.1093/mnras/stt2174}, \href
  {https://ui.adsabs.harvard.edu/abs/2014MNRAS.437.3787C} {437, 3787}

\bibitem[\protect\citeauthoryear{{Cooper} et~al.,}{{Cooper}
  et~al.}{2010}]{cooper10}
{Cooper} A.~P.,  et~al., 2010, \mn@doi [\mnras]
  {10.1111/j.1365-2966.2010.16740.x}, \href
  {https://ui.adsabs.harvard.edu/abs/2010MNRAS.406..744C} {406, 744}

\bibitem[\protect\citeauthoryear{{Cooper}, {Gao}, {Guo}, {Frenk}, {Jenkins},
  {Springel}  \& {White}}{{Cooper} et~al.}{2015}]{cooper15}
{Cooper} A.~P.,  {Gao} L.,  {Guo} Q.,  {Frenk} C.~S.,  {Jenkins} A.,
  {Springel} V.,   {White} S.~D.~M.,  2015, \mn@doi [\mnras]
  {10.1093/mnras/stv1042}, \href
  {https://ui.adsabs.harvard.edu/abs/2015MNRAS.451.2703C} {451, 2703}

\bibitem[\protect\citeauthoryear{{Crain} et~al.,}{{Crain}
  et~al.}{2015}]{crain15}
{Crain} R.~A.,  et~al., 2015, \mn@doi [\mnras] {10.1093/mnras/stv725}, \href
  {https://ui.adsabs.harvard.edu/abs/2015MNRAS.450.1937C} {450, 1937}

\bibitem[\protect\citeauthoryear{{Davis}, {Efstathiou}, {Frenk}  \&
  {White}}{{Davis} et~al.}{1985}]{davis85}
{Davis} M.,  {Efstathiou} G.,  {Frenk} C.~S.,   {White} S.~D.~M.,  1985,
  \mn@doi [\apj] {10.1086/163168}, \href
  {https://ui.adsabs.harvard.edu/abs/1985ApJ...292..371D} {292, 371}

\bibitem[\protect\citeauthoryear{{DeMaio}, {Gonzalez}, {Zabludoff}, {Zaritsky},
  {Connor}, {Donahue}  \& {Mulchaey}}{{DeMaio} et~al.}{2018}]{demaio18}
{DeMaio} T.,  {Gonzalez} A.~H.,  {Zabludoff} A.,  {Zaritsky} D.,  {Connor} T.,
  {Donahue} M.,   {Mulchaey} J.~S.,  2018, \mn@doi [\mnras]
  {10.1093/mnras/stx2946}, \href
  {https://ui.adsabs.harvard.edu/abs/2018MNRAS.474.3009D} {474, 3009}

\bibitem[\protect\citeauthoryear{{Deason}, {Belokurov}  \& {Weisz}}{{Deason}
  et~al.}{2015}]{deason15}
{Deason} A.~J.,  {Belokurov} V.,   {Weisz} D.~R.,  2015, \mn@doi [\mnras]
  {10.1093/mnrasl/slv001}, \href
  {https://ui.adsabs.harvard.edu/abs/2015MNRAS.448L..77D} {448, L77}

\bibitem[\protect\citeauthoryear{{Deason}, {Mao}  \& {Wechsler}}{{Deason}
  et~al.}{2016}]{deason16}
{Deason} A.~J.,  {Mao} Y.-Y.,   {Wechsler} R.~H.,  2016, \mn@doi [\apj]
  {10.3847/0004-637X/821/1/5}, \href
  {https://ui.adsabs.harvard.edu/abs/2016ApJ...821....5D} {821, 5}

\bibitem[\protect\citeauthoryear{{Deason}, {Fattahi}, {Frenk}, {Grand },
  {Oman}, {Garrison-Kimmel}, {Simpson}  \& {Navarro}}{{Deason}
  et~al.}{2020}]{deason20}
{Deason} A.~J.,  {Fattahi} A.,  {Frenk} C.~S.,  {Grand } R. J.~J.,  {Oman}
  K.~A.,  {Garrison-Kimmel} S.,  {Simpson} C.~M.,   {Navarro} J.~F.,  2020,
  \mn@doi [\mnras] {10.1093/mnras/staa1711}, \href
  {https://ui.adsabs.harvard.edu/abs/2020MNRAS.496.3929D} {496, 3929}

\bibitem[\protect\citeauthoryear{{Diemer}}{{Diemer}}{2020}]{diemer20}
{Diemer} B.,  2020, arXiv e-prints, \href
  {https://ui.adsabs.harvard.edu/abs/2020arXiv200709149D} {p. arXiv:2007.09149}

\bibitem[\protect\citeauthoryear{{Diemer} \& {Kravtsov}}{{Diemer} \&
  {Kravtsov}}{2014}]{diemer14}
{Diemer} B.,  {Kravtsov} A.~V.,  2014, \mn@doi [\apj]
  {10.1088/0004-637X/789/1/1}, \href
  {https://ui.adsabs.harvard.edu/abs/2014ApJ...789....1D} {789, 1}

\bibitem[\protect\citeauthoryear{{Diemer}, {Mansfield}, {Kravtsov}  \&
  {More}}{{Diemer} et~al.}{2017}]{diemer17}
{Diemer} B.,  {Mansfield} P.,  {Kravtsov} A.~V.,   {More} S.,  2017, \mn@doi
  [\apj] {10.3847/1538-4357/aa79ab}, \href
  {https://ui.adsabs.harvard.edu/abs/2017ApJ...843..140D} {843, 140}

\bibitem[\protect\citeauthoryear{{Fakhouri} \& {Ma}}{{Fakhouri} \&
  {Ma}}{2010}]{fakhouri10}
{Fakhouri} O.,  {Ma} C.-P.,  2010, \mn@doi [\mnras]
  {10.1111/j.1365-2966.2009.15844.x}, \href
  {https://ui.adsabs.harvard.edu/abs/2010MNRAS.401.2245F} {401, 2245}

\bibitem[\protect\citeauthoryear{{Fielder}, {Mao}, {Zentner}, {Newman}, {Wu}
  \& {Wechsler}}{{Fielder} et~al.}{2020}]{fielder20}
{Fielder} C.~E.,  {Mao} Y.-Y.,  {Zentner} A.~R.,  {Newman} J.~A.,  {Wu} H.-Y.,
   {Wechsler} R.~H.,  2020, \mn@doi [\mnras] {10.1093/mnras/staa2851}, \href
  {https://ui.adsabs.harvard.edu/abs/2020MNRAS.499.2426F} {499, 2426}

\bibitem[\protect\citeauthoryear{{Gao}, {Frenk}, {Boylan-Kolchin}, {Jenkins},
  {Springel}  \& {White}}{{Gao} et~al.}{2011}]{gao11}
{Gao} L.,  {Frenk} C.~S.,  {Boylan-Kolchin} M.,  {Jenkins} A.,  {Springel} V.,
   {White} S.~D.~M.,  2011, \mn@doi [\mnras]
  {10.1111/j.1365-2966.2010.17601.x}, \href
  {https://ui.adsabs.harvard.edu/abs/2011MNRAS.410.2309G} {410, 2309}

\bibitem[\protect\citeauthoryear{{Genel}, {Bouch{\'e}}, {Naab}, {Sternberg}  \&
  {Genzel}}{{Genel} et~al.}{2010}]{genel10}
{Genel} S.,  {Bouch{\'e}} N.,  {Naab} T.,  {Sternberg} A.,   {Genzel} R.,
  2010, \mn@doi [\apj] {10.1088/0004-637X/719/1/229}, \href
  {https://ui.adsabs.harvard.edu/abs/2010ApJ...719..229G} {719, 229}

\bibitem[\protect\citeauthoryear{{Ivezi{\'c}} et~al.,}{{Ivezi{\'c}}
  et~al.}{2019}]{lsst}
{Ivezi{\'c}} {\v Z}.,  et~al., 2019, \mn@doi [\apj] {10.3847/1538-4357/ab042c},
  \href {http://adsabs.harvard.edu/abs/2019ApJ...873..111I} {873, 111}

\bibitem[\protect\citeauthoryear{{Laureijs} et~al.,}{{Laureijs}
  et~al.}{2011}]{euclid}
{Laureijs} R.,  et~al., 2011, arXiv e-prints, \href
  {https://ui.adsabs.harvard.edu/abs/2011arXiv1110.3193L} {p. arXiv:1110.3193}

\bibitem[\protect\citeauthoryear{{Mansfield}, {Kravtsov}  \&
  {Diemer}}{{Mansfield} et~al.}{2017}]{mansfield17}
{Mansfield} P.,  {Kravtsov} A.~V.,   {Diemer} B.,  2017, \mn@doi [\apj]
  {10.3847/1538-4357/aa7047}, \href
  {https://ui.adsabs.harvard.edu/abs/2017ApJ...841...34M} {841, 34}

\bibitem[\protect\citeauthoryear{{McConnachie}}{{McConnachie}}{2012}]{mcconnachie12}
{McConnachie} A.~W.,  2012, \mn@doi [\aj] {10.1088/0004-6256/144/1/4}, \href
  {https://ui.adsabs.harvard.edu/abs/2012AJ....144....4M} {144, 4}

\bibitem[\protect\citeauthoryear{{Mihos}}{{Mihos}}{2016}]{mihos16}
{Mihos} J.~C.,  2016, in {Bragaglia} A.,  {Arnaboldi} M.,  {Rejkuba} M.,
  {Romano} D.,  eds,  IAU Symposium Vol. 317, The General Assembly of Galaxy
  Halos: Structure, Origin and Evolution. pp 27--34 (\mn@eprint {arXiv}
  {1510.01929}), \mn@doi{10.1017/S1743921315006857}

\bibitem[\protect\citeauthoryear{{Montes}}{{Montes}}{2019}]{montes19_rev}
{Montes} M.,  2019, arXiv e-prints, \href
  {https://ui.adsabs.harvard.edu/abs/2019arXiv191201616M} {p. arXiv:1912.01616}

\bibitem[\protect\citeauthoryear{{Montes} \& {Trujillo}}{{Montes} \&
  {Trujillo}}{2014}]{montes14}
{Montes} M.,  {Trujillo} I.,  2014, \mn@doi [\apj]
  {10.1088/0004-637X/794/2/137}, \href
  {https://ui.adsabs.harvard.edu/abs/2014ApJ...794..137M} {794, 137}

\bibitem[\protect\citeauthoryear{{Montes} \& {Trujillo}}{{Montes} \&
  {Trujillo}}{2018}]{montes18}
{Montes} M.,  {Trujillo} I.,  2018, \mn@doi [\mnras] {10.1093/mnras/stx2847},
  \href {https://ui.adsabs.harvard.edu/abs/2018MNRAS.474..917M} {474, 917}

\bibitem[\protect\citeauthoryear{{Montes} \& {Trujillo}}{{Montes} \&
  {Trujillo}}{2019}]{montes19}
{Montes} M.,  {Trujillo} I.,  2019, \mn@doi [\mnras] {10.1093/mnras/sty2858},
  \href {https://ui.adsabs.harvard.edu/abs/2019MNRAS.482.2838M} {482, 2838}

\bibitem[\protect\citeauthoryear{{More}, {Diemer}  \& {Kravtsov}}{{More}
  et~al.}{2015}]{more15}
{More} S.,  {Diemer} B.,   {Kravtsov} A.~V.,  2015, \mn@doi [\apj]
  {10.1088/0004-637X/810/1/36}, \href
  {https://ui.adsabs.harvard.edu/abs/2015ApJ...810...36M} {810, 36}

\bibitem[\protect\citeauthoryear{{More} et~al.,}{{More} et~al.}{2016}]{more16}
{More} S.,  et~al., 2016, \mn@doi [\apj] {10.3847/0004-637X/825/1/39}, \href
  {https://ui.adsabs.harvard.edu/abs/2016ApJ...825...39M} {825, 39}

\bibitem[\protect\citeauthoryear{{Moster}, {Somerville}, {Maulbetsch}, {van den
  Bosch}, {Macci{\`o}}, {Naab}  \& {Oser}}{{Moster} et~al.}{2010}]{moster10}
{Moster} B.~P.,  {Somerville} R.~S.,  {Maulbetsch} C.,  {van den Bosch} F.~C.,
  {Macci{\`o}} A.~V.,  {Naab} T.,   {Oser} L.,  2010, \mn@doi [\apj]
  {10.1088/0004-637X/710/2/903}, \href
  {https://ui.adsabs.harvard.edu/abs/2010ApJ...710..903M} {710, 903}

\bibitem[\protect\citeauthoryear{{Murante} et~al.,}{{Murante}
  et~al.}{2004}]{murante04}
{Murante} G.,  et~al., 2004, \mn@doi [\apjl] {10.1086/421348}, \href
  {https://ui.adsabs.harvard.edu/abs/2004ApJ...607L..83M} {607, L83}

\bibitem[\protect\citeauthoryear{{Murata}, {Sunayama}, {Oguri}, {More},
  {Nishizawa}, {Nishimichi}  \& {Osato}}{{Murata} et~al.}{2020}]{murata20}
{Murata} R.,  {Sunayama} T.,  {Oguri} M.,  {More} S.,  {Nishizawa} A.~J.,
  {Nishimichi} T.,   {Osato} K.,  2020, \mn@doi [\pasj] {10.1093/pasj/psaa041},
  \href {https://ui.adsabs.harvard.edu/abs/2020PASJ...72...64M} {72, 64}

\bibitem[\protect\citeauthoryear{{Pe{\~n}arrubia}, {Navarro}  \&
  {McConnachie}}{{Pe{\~n}arrubia} et~al.}{2008}]{penarrubia08}
{Pe{\~n}arrubia} J.,  {Navarro} J.~F.,   {McConnachie} A.~W.,  2008, \mn@doi
  [\apj] {10.1086/523686}, \href
  {https://ui.adsabs.harvard.edu/abs/2008ApJ...673..226P} {673, 226}

\bibitem[\protect\citeauthoryear{{Pearce}, {Kay}, {Barnes}, {Bahe}  \&
  {Bower}}{{Pearce} et~al.}{2020}]{pearce20}
{Pearce} F.~A.,  {Kay} S.~T.,  {Barnes} D.~J.,  {Bahe} Y.~M.,   {Bower} R.~G.,
  2020, arXiv e-prints, \href
  {https://ui.adsabs.harvard.edu/abs/2020arXiv200512391P} {p. arXiv:2005.12391}

\bibitem[\protect\citeauthoryear{{Pillepich} et~al.,}{{Pillepich}
  et~al.}{2014}]{pillepich14}
{Pillepich} A.,  et~al., 2014, \mn@doi [\mnras] {10.1093/mnras/stu1408}, \href
  {https://ui.adsabs.harvard.edu/abs/2014MNRAS.444..237P} {444, 237}

\bibitem[\protect\citeauthoryear{{Pillepich} et~al.,}{{Pillepich}
  et~al.}{2018}]{pillepich18}
{Pillepich} A.,  et~al., 2018, \mn@doi [\mnras] {10.1093/mnras/stx3112}, \href
  {https://ui.adsabs.harvard.edu/abs/2018MNRAS.475..648P} {475, 648}

\bibitem[\protect\citeauthoryear{{Planck Collaboration} et~al.,}{{Planck
  Collaboration} et~al.}{2014}]{planck14}
{Planck Collaboration} et~al., 2014, \mn@doi [\aap]
  {10.1051/0004-6361/201321591}, \href
  {https://ui.adsabs.harvard.edu/abs/2014A&A...571A..16P} {571, A16}

\bibitem[\protect\citeauthoryear{{Puchwein}, {Springel}, {Sijacki}  \&
  {Dolag}}{{Puchwein} et~al.}{2010}]{puchwein10}
{Puchwein} E.,  {Springel} V.,  {Sijacki} D.,   {Dolag} K.,  2010, \mn@doi
  [\mnras] {10.1111/j.1365-2966.2010.16786.x}, \href
  {https://ui.adsabs.harvard.edu/abs/2010MNRAS.406..936P} {406, 936}

\bibitem[\protect\citeauthoryear{{Purcell}, {Bullock}  \& {Zentner}}{{Purcell}
  et~al.}{2007}]{purcell07}
{Purcell} C.~W.,  {Bullock} J.~S.,   {Zentner} A.~R.,  2007, \mn@doi [\apj]
  {10.1086/519787}, \href
  {https://ui.adsabs.harvard.edu/abs/2007ApJ...666...20P} {666, 20}

\bibitem[\protect\citeauthoryear{{Robotham} et~al.,}{{Robotham}
  et~al.}{2011}]{robotham11}
{Robotham} A.~S.~G.,  et~al., 2011, \mn@doi [\mnras]
  {10.1111/j.1365-2966.2011.19217.x}, \href
  {https://ui.adsabs.harvard.edu/abs/2011MNRAS.416.2640R} {416, 2640}

\bibitem[\protect\citeauthoryear{{Rudick}, {Mihos}  \& {McBride}}{{Rudick}
  et~al.}{2011}]{rudick11}
{Rudick} C.~S.,  {Mihos} J.~C.,   {McBride} C.~K.,  2011, \mn@doi [\apj]
  {10.1088/0004-637X/732/1/48}, \href
  {https://ui.adsabs.harvard.edu/abs/2011ApJ...732...48R} {732, 48}

\bibitem[\protect\citeauthoryear{{Sampaio-Santos} et~al.,}{{Sampaio-Santos}
  et~al.}{2020}]{sampaio20}
{Sampaio-Santos} H.,  et~al., 2020, arXiv e-prints, \href
  {https://ui.adsabs.harvard.edu/abs/2020arXiv200512275S} {p. arXiv:2005.12275}

\bibitem[\protect\citeauthoryear{{Savitzky} \& {Golay}}{{Savitzky} \&
  {Golay}}{1964}]{savitzky64}
{Savitzky} A.,  {Golay} M.~J.~E.,  1964, Analytical Chemistry, \href
  {https://ui.adsabs.harvard.edu/abs/1964AnaCh..36.1627S} {36, 1627}

\bibitem[\protect\citeauthoryear{{Schaller} et~al.,}{{Schaller}
  et~al.}{2015}]{schaller15}
{Schaller} M.,  et~al., 2015, \mn@doi [\mnras] {10.1093/mnras/stv1067}, \href
  {https://ui.adsabs.harvard.edu/abs/2015MNRAS.451.1247S} {451, 1247}

\bibitem[\protect\citeauthoryear{{Schaye} et~al.,}{{Schaye}
  et~al.}{2015}]{schaye15}
{Schaye} J.,  et~al., 2015, \mn@doi [\mnras] {10.1093/mnras/stu2058}, \href
  {https://ui.adsabs.harvard.edu/abs/2015MNRAS.446..521S} {446, 521}

\bibitem[\protect\citeauthoryear{{Shin} et~al.,}{{Shin} et~al.}{2019}]{shin19}
{Shin} T.,  et~al., 2019, \mn@doi [\mnras] {10.1093/mnras/stz1434}, \href
  {https://ui.adsabs.harvard.edu/abs/2019MNRAS.487.2900S} {487, 2900}

\bibitem[\protect\citeauthoryear{{Spergel} et~al.,}{{Spergel}
  et~al.}{2015}]{wfirst}
{Spergel} D.,  et~al., 2015, arXiv e-prints, \href
  {https://ui.adsabs.harvard.edu/abs/2015arXiv150303757S} {p. arXiv:1503.03757}

\bibitem[\protect\citeauthoryear{{Springel}, {White}, {Tormen}  \&
  {Kauffmann}}{{Springel} et~al.}{2001}]{springel01}
{Springel} V.,  {White} S. D.~M.,  {Tormen} G.,   {Kauffmann} G.,  2001,
  \mn@doi [\mnras] {10.1046/j.1365-8711.2001.04912.x}, \href
  {https://ui.adsabs.harvard.edu/abs/2001MNRAS.328..726S} {328, 726}

\bibitem[\protect\citeauthoryear{{Steinhardt} et~al.,}{{Steinhardt}
  et~al.}{2020}]{buffalo}
{Steinhardt} C.~L.,  et~al., 2020, \mn@doi [\apjs] {10.3847/1538-4365/ab75ed},
  \href {https://ui.adsabs.harvard.edu/abs/2020ApJS..247...64S} {247, 64}

\bibitem[\protect\citeauthoryear{{Tam} et~al.,}{{Tam} et~al.}{2020}]{tam20}
{Tam} S.-I.,  et~al., 2020, \mn@doi [\mnras] {10.1093/mnras/staa1828}, \href
  {https://ui.adsabs.harvard.edu/abs/2020MNRAS.496.4032T} {496, 4032}

\bibitem[\protect\citeauthoryear{{Turk}, {Smith}, {Oishi}, {Skory}, {Skillman},
  {Abel}  \& {Norman}}{{Turk} et~al.}{2011}]{yt}
{Turk} M.~J.,  {Smith} B.~D.,  {Oishi} J.~S.,  {Skory} S.,  {Skillman} S.~W.,
  {Abel} T.,   {Norman} M.~L.,  2011, \mn@doi [\apjs]
  {10.1088/0067-0049/192/1/9}, \href
  {https://ui.adsabs.harvard.edu/abs/2011ApJS..192....9T} {192, 9}

\bibitem[\protect\citeauthoryear{{Wang} et~al.,}{{Wang} et~al.}{2011}]{wang11}
{Wang} J.,  et~al., 2011, \mn@doi [\mnras] {10.1111/j.1365-2966.2011.18220.x},
  \href {https://ui.adsabs.harvard.edu/abs/2011MNRAS.413.1373W} {413, 1373}

\bibitem[\protect\citeauthoryear{{Watson}, {Berlind}  \& {Zentner}}{{Watson}
  et~al.}{2012}]{watson12}
{Watson} D.~F.,  {Berlind} A.~A.,   {Zentner} A.~R.,  2012, \mn@doi [\apj]
  {10.1088/0004-637X/754/2/90}, \href
  {https://ui.adsabs.harvard.edu/abs/2012ApJ...754...90W} {754, 90}

\bibitem[\protect\citeauthoryear{{Xhakaj}, {Diemer}, {Leauthaud}, {Wasserman},
  {Huang}, {Luo}, {Adhikari}  \& {Singh}}{{Xhakaj} et~al.}{2020}]{xhakaj20}
{Xhakaj} E.,  {Diemer} B.,  {Leauthaud} A.,  {Wasserman} A.,  {Huang} S.,
  {Luo} Y.,  {Adhikari} S.,   {Singh} S.,  2020, \mn@doi [\mnras]
  {10.1093/mnras/staa3046}, \href
  {https://ui.adsabs.harvard.edu/abs/2020MNRAS.499.3534X} {499, 3534}

\bibitem[\protect\citeauthoryear{{Yang}, {Mo}, {van den Bosch}  \&
  {Jing}}{{Yang} et~al.}{2005}]{yang05}
{Yang} X.,  {Mo} H.~J.,  {van den Bosch} F.~C.,   {Jing} Y.~P.,  2005, \mn@doi
  [\mnras] {10.1111/j.1365-2966.2005.08560.x}, \href
  {https://ui.adsabs.harvard.edu/abs/2005MNRAS.356.1293Y} {356, 1293}

\bibitem[\protect\citeauthoryear{{Zhang} et~al.,}{{Zhang}
  et~al.}{2019}]{zhang19}
{Zhang} Y.,  et~al., 2019, \mn@doi [\apj] {10.3847/1538-4357/ab0dfd}, \href
  {https://ui.adsabs.harvard.edu/abs/2019ApJ...874..165Z} {874, 165}

\bibitem[\protect\citeauthoryear{{Zibetti}, {White}, {Schneider}  \&
  {Brinkmann}}{{Zibetti} et~al.}{2005}]{zibetti05}
{Zibetti} S.,  {White} S. D.~M.,  {Schneider} D.~P.,   {Brinkmann} J.,  2005,
  \mn@doi [\mnras] {10.1111/j.1365-2966.2005.08817.x}, \href
  {https://ui.adsabs.harvard.edu/abs/2005MNRAS.358..949Z} {358, 949}

\bibitem[\protect\citeauthoryear{{Z{\"u}rcher} \& {More}}{{Z{\"u}rcher} \&
  {More}}{2019}]{zurcher19}
{Z{\"u}rcher} D.,  {More} S.,  2019, \mn@doi [\apj] {10.3847/1538-4357/ab08e8},
  \href {https://ui.adsabs.harvard.edu/abs/2019ApJ...874..184Z} {874, 184}

\bibitem[\protect\citeauthoryear{{van den Bosch}, {Yang}, {Mo}  \&
  {Norberg}}{{van den Bosch} et~al.}{2005}]{vandenbosch05}
{van den Bosch} F.~C.,  {Yang} X.,  {Mo} H.~J.,   {Norberg} P.,  2005, \mn@doi
  [\mnras] {10.1111/j.1365-2966.2004.08407.x}, \href
  {https://ui.adsabs.harvard.edu/abs/2005MNRAS.356.1233V} {356, 1233}

\makeatother
\end{thebibliography}

\label{lastpage}
\end{document}